\documentclass[12pt,preprint]{aastex}

\shorttitle{LISM III: Temperature and Turbulence}
\shortauthors{}

\begin{document}

\newcommand{\php}[0]{\phantom{--}}
\newcommand{\kms}[0]{km~s$^{-1}$}

\title{The Structure of the Local Interstellar Medium. III. Temperature and 
Turbulence\label{tt}\footnote{Based on observations made with the NASA/ESA 
Hubble Space Telescope, obtained from the Data Archive at the Space Telescope 
Science Institute, which is operated by the Association of Universities for 
Research in Astronomy, Inc., under NASA contract NAS AR-09525.01A. These 
observations are associated with program \#9525.}}

\author{Seth Redfield\altaffilmark{2}, \& Jeffrey L. Linsky}
\affil{JILA, University of Colorado and NIST, Boulder, CO 80309-0440}
\altaffiltext{2}{Currently a Harlan J. Smith Postdoctoral Fellow at McDonald 
Observatory, Austin, TX 78712-0259; sredfield@astro.as.utexas.edu}

\begin{abstract}

We present 50 individual measurements of the 
gas temperature and turbulent 
velocity in the local interstellar medium (LISM) within 100\,pc.  By 
comparing the 
absorption line widths of many ions with different atomic 
masses, we can 
satisfactorily 
discriminate between the two dominant broadening mechanisms, 
thermal broadening, and macroscopic nonthermal, or turbulent, broadening.  We 
find that the successful use of this technique requires a measurement of a 
light ion, such as \ion{D}{1}, and an ion at least as heavy as \ion{Mg}{2}.  
However, observations of more lines provides an important consistency check 
and can also improve the precision and accuracy of the measurement.  
Temperature and turbulent velocity measurements are vital to understanding 
the physical properties of the gas in our local environment, and can 
provide insight into the three-dimensional morphological structure of the 
LISM.  The weighted mean 
gas temperature in the LISM warm clouds is 6680\,K and 
the dispersion about the mean is 1490\,K.  
The weighted mean 
turbulent velocity is 2.24\,km\,s$^{-1}$ and the dispersion about the 
mean is 1.03\,km\,s$^{-1}$.  The ratio of 
the mean thermal pressure to the mean turbulent 
pressure is $P_{\rm T}/P_{\xi}\,\sim\,26$.  Turbulent pressure in LISM clouds 
cannot explain the difference in the apparent pressure imbalance between warm 
LISM clouds and the surrounding hot gas of the Local Bubble.  Pressure 
equilibrium 
among the warm clouds may be the source of a 
moderately negative correlation between 
temperature and turbulent velocity in these clouds.  However, significant 
variations 
in temperature and turbulent velocity are observed.
The turbulent motions in the 
warm partially ionized clouds of the LISM are 
definitely subsonic, and the 
weighted mean turbulent Mach number for clouds in the LISM is 
0.19 with a dispersion of 0.11.  These measurements provide important constraints on 
models of the evolution and origin of warm partially ionized clouds in our 
local environment.

\end{abstract}


\section{Introduction}

Understanding the physical state of the interstellar medium (ISM) is a 
fundamental area of research in astrophysics.  Theoretical studies of the 
phases of the ISM have produced classic papers in the astrophysical 
literature 
\citep[e.g.,][]{field69,mckee77}, the details of which are still being 
discussed, analyzed, and refined to this day 
\citep{cox95,wolfire95a,wolfire95b,heiles01}.  
These classical theoretical models predict the stable 
coexistence in pressure equilibrium of various phases of interstellar matter:
the cold neutral medium (CNM), 
the warm neutral medium (WNM),
the warm ionized medium (WIM),
and the hot ionized medium (HIM).  Detailed observations are 
required to test the theory and provide direction for future modifications.  

The heliosphere is surrounded by warm 
partially ionized material \citep{red00}
in the local interstellar medium (LISM), which has properties intermediate 
between those of the WNM and WIM theoretical models. 
The LISM is embedded in a larger 
cavity, the Local Bubble, presumably filled with hot rarefied gas.  Very 
little cold material, or CNM, resides in the LISM, as is shown by the lack of 
\ion{Na}{1} absorption towards nearby targets \citep{lall03,sf99,welty96,welsh94}.  
However, observations of this 
cold material are made from low ionization 
absorption line measurements \citep{jenkins01}, and from absorption and 
emission in the 21\,cm hyperfine structure line of \ion{H}{1} \citep{ch01}.  The 
HIM that presumably pervades the Local Bubble, was identified by the presence 
of soft X-ray emission by \citet{snowden98}. 
The {\it Cosmic Hot 
Interstellar Plasma Spectrometer} ({\it CHIPS}) instrument, which was 
launched on 
2003 January 12, should provide important new observations of 
this rarefied gas \citep{mh99}.  
 
Observations of the warm partially ionized 
gas are 
vital to unlocking the physical structure of the ISM.  Distant WNM 
environments can be probed by analysis of the 21\,cm 
hyperfine structure line of \ion{H}{1}, where the WNM is detected in emission and the CNM in absorption.  
However, only upper limits of the temperature 
can be inferred because the contribution of nonthermal, or turbulent, 
broadening cannot be measured 
separately \citep{ch01}.  Ultraviolet (UV) spectra of 
absorption lines can be used to make accurate temperature and turbulent 
velocity measurements of 
gas in the LISM by comparing the observed line widths of 
ions with different atomic masses.  However, for distant lines of sight, the 
absorption features are typically severely blended, and most information 
regarding the influences of thermal and turbulent broadening mechanisms on 
the line widths is lost.  Only crude estimates of the temperature can be made 
from the detection or nondetection of various ionization stages of observed 
elements \citep{welty02}.  Therefore, nearby targets, and the intervening 
warm partially ionized 
gas, represent a unique opportunity to analyze 
unsaturated absorption lines 
with simple and well-characterized component structures 
so as to 
measure accurately the temperature and turbulence 
along lines of sight to these targets.

After more than a decade of high resolution UV spectroscopy with the {\it 
Hubble Space Telescope} ({\it HST}), we are beginning to accumulate a 
critical number of observations of the LISM.  Collections of observations 
presented by \citet{red02,red03a}, and researchers referenced therein, 
provide an opportunity to make fundamental physical measurements of the 
velocity, temperature, and turbulent velocity 
in the gas along the line of sight.  
Researchers investigating the 
LISM absorption lines towards nearby stars 
typically calculate a temperature and 
turbulent velocity
for a given line of sight, but do not
analyze the results in the context of the LISM 
as a whole.  We present here the first systematic survey of temperature and 
turbulent velocity measurements for the 
LISM gas within 100 pc of the Sun.

\section{Calculating the Temperature and Turbulent Velocity \label{calctxi}}

By measuring the LISM absorption line widths of many ions, one can 
disentangle the 
contributions of thermal and turbulent 
broadening.  The standard equation that relates the absorption line width, or 
Doppler parameter ($b$), and the temperature ($T$) and turbulent velocity 
($\xi$) of the absorbing gas is 
\begin{equation}
b^2 = \frac{2kT}{m}+\xi^2 = 0.016629\frac{T}{A}+\xi^2 ,
\label{tt_eq1}
\end{equation}
where $k$ is Boltzmann's constant, $m$ is the mass of the ion, and $A$ is the 
atomic weight of the ion in atomic mass units.  Although only two 
line width measurements are required to solve 
Equation~\ref{tt_eq1} for the temperature 
and turbulence, an accurate solution requires that the two ions have very 
different masses.  Ideally, one would like to 
compare line widths of the two ions with 
the greatest mass differences
that are observed in the LISM, hydrogen and iron.  Iron is an excellent 
candidate because there are a number of resonance lines in the 
2000-2600\,\AA\ region that span a range of optical depths, permitting precise 
measurements of unsaturated absorption lines.  In addition, due to its large 
mass, thermal broadening is typically negligible and the line is 
intrinsically narrow, which 
improves the precision of the line fitting 
and facilitates the identification of individual velocity components.  
\citet{red02} presented an inventory of all measured \ion{Fe}{2} LISM 
absorption line measurements toward stars within 100\,pc.  

The light ions, however, pose more of a problem.  Although the strong 
Ly-$\alpha$ line at 1215\,\AA\ is available for measuring the \ion{H}{1} absorption 
line width, 
the line is heavily saturated and very broad, which makes estimating 
the stellar continuum difficult.  This line can also be broadened by 
heliospheric and astrospheric absorption components 
\citep[e.g.,][]{lin96,wood00}.  
Deuterium is the next lightest ion, only twice as heavy as hydrogen.  The 
\ion{D}{1} absorption line is offset by roughly $-82\,$km\,s$^{-1}$ 
from the \ion{H}{1} Ly-$\alpha$ line.  For short lines of sight the \ion{D}{1} 
line is typically unsaturated.  Since 
the \ion{D}{1} line usually lies on the extended wing of 
the broad stellar Ly-$\alpha$ emission line, it is relatively straightforward 
to interpolate 
the background continuum.  \ion{D}{1} 
is the subject of many papers and 
has been analyzed by many groups.  \citet{red03a} bring together the full 
collection of \ion{D}{1} Ly-$\alpha$ observations towards stars within 
100\,pc with known velocity component structures, including many new measurements.  

In Figures~\ref{tt_fig1} through \ref{tt_fig13}, we present our measurements 
of the temperature and turbulent velocity for all 50 
observed LISM components along 29 lines of sight to stars located 
within 100\,pc that have absorption line width measurements of \ion{D}{1} and 
an ion at least as heavy as \ion{Mg}{2}.  Sources for all 
of the Doppler width 
measurements used in Figures~\ref{tt_fig1}-\ref{tt_fig13} can be found in 
\citet{red02} and \citet{red03a}.  For each 
velocity component along a particular 
sightline, there are two plots to visualize the temperature and turbulent 
velocity measurement.  On the left side is a color coded plot of temperature 
versus turbulent velocity.  For a given measured Doppler parameter of a 
particular ion, Equation~\ref{tt_eq1} defines a curve in the 
temperature-turbulent velocity plane.  The accompanying pair of dashed lines 
indicate the $\pm\,1\sigma$ error bars around the best fit curve
(solid line).  The black 
``$\times$'' symbol denotes the best fit value of the temperature and 
turbulent velocity 
based on the widths of the deuterium line 
and all other line widths measured from high resolution 
echelle (STIS E140H and E230H and GHRS Ech-A and Ech-B) spectra.
The surrounding black contours enclose the 1$\sigma$ 
and 2$\sigma$ ranges for both values.  The $1\sigma$ and 2$\sigma$ error 
contours were determined by calculating the $\Delta\chi$ appropriate for the 
68.27\% and 95.45\% confidence level, respectively, as described in 
\citet{bev92}.  Except for deuterium, all the other absorption lines are not fully 
resolved at moderate resolution.  Therefore, systematic errors, such as, 
inaccurate instrumental line spread functions, unresolved blends, and 
saturated lines, can potentially dominate the moderate resolution results.  
For this reason, the fits were made 
only to the deuterium measurement and all 
additional high resolution measurements.  The moderate resolution data 
are plotted as consistency checks.  Two sightlines, HR\,1099 and $\beta$\,Cet, analyzed by \citet{piskunov97}, use the \ion{H}{1} Doppler width as opposed to \ion{D}{1}.  Although they did not provide an independent \ion{D}{1} line width measurement, their \ion{H}{1} Doppler width was forced to agree with the \ion{D}{1} line.  Therefore, despite the difficulty of analyzing the \ion{H}{1} absorption line, these are likely to be reasonable estimates of the \ion{H}{1} Doppler parameter.

The plots on the right show another visualization 
for determining 
the temperature and turbulent velocity.  Here, the observed Doppler parameter 
for each ion is plotted versus its atomic 
mass.  The red and blue symbols 
represent the observations that were taken with high resolution and moderate 
resolution, respectively.  The solid line indicates the best fit curve for 
the temperature and turbulence, as defined by Equation~\ref{tt_eq1}.  The 
shaded region encompasses all fits within the $1\sigma$ contour shown in the 
plot on the left.  

Both plots clearly demonstrate the need for using both the light ion, 
deuterium, and preferably the heaviest ion, iron, for calculating the 
temperature and turbulent velocity.  The deuterium curve in the plots on the 
left is 
nearly vertical, whereas the curve for iron is practically 
horizontal, indicating that for deuterium thermal broadening dominates, and 
the line width is relatively insensitive to turbulent broadening.  The 
reverse is true for iron.  In the plots on the right, \ion{D}{1} lies on the 
steep part of the curve, while practically all the remaining ions are on the 
flat part of the curve, where turbulent broadening dominates, and the line 
widths are independent of atomic 
mass and thermal broadening.  Deuterium is 
significantly broader than all 
of the other ions and is the only serious 
constraint on the temperature of the gas.  Even the next lightest ion 
observed in the LISM, \ion{C}{2}, is already 6 times the 
mass of deuterium, 
and predominantly horizontal in the plots on the left and on the flat part of 
the curve in the plots on the right.  

Although only two measurements are required to solve Equation~\ref{tt_eq1}, 
the intermediate mass ions 
provide important consistency checks against 
potentially unidentified systematic errors in the heavy and light ion 
measurements.  Often, the lines of \ion{C}{2} and \ion{O}{1} have large 
errors in the Doppler parameter measurement because they are highly 
saturated.  HZ~43 in Figure~\ref{tt_fig10} provides an excellent example of 
agreement among all of the measured ions.  Only a handful of components show 
significant disagreement among the observed ions.  In particular, Altair 
(Figures~\ref{tt_fig3} and \ref{tt_fig4}) shows 
possible discrepancies in the 
observed \ion{C}{2} Doppler width.  This is most likely due to the \ion{C}{2} 
absorption line being highly saturated.  Low signal-to-noise observations or multiple closely spaced components can also make a precise Doppler width measurement difficult.  Otherwise, this 
technique offers a unique opportunity to measure accurately the temperature 
and turbulent velocity in distinct collections of gas in the LISM.  

Table~\ref{tt_table1} lists the best fit temperature and turbulent velocity 
for all components shown in Figures~\ref{tt_fig1} through \ref{tt_fig13}.  
The $\pm\,1\sigma$ error bars were determined by fixing each parameter in 
turn, and calculating the $\Delta\chi$ appropriate for the 68.27\% confidence 
level, as described in \citet{bev92}.  
To determine the best fit temperatures and turbulent velocities for each 
velocity component, we weighted the
observed Doppler parameters 
inversely to their measurement errors.  We also include, for each 
component, the list of ions for which a Doppler width was included in the 
fit.  Two is the minimum number of ions required for this technique, but our 
fits typically include Doppler widths from $\sim$\,5 different 
ions, and in some cases as many as eight.  Also included in 
Table~\ref{tt_table1}, are the galactic coordinates and distance to the background star, and the weighted mean velocity measurements for the 
observed LISM absorption components.  

\section{Temperature and Turbulent Velocity Distribution in the LISM}

The distributions of temperatures and turbulent velocities in the LISM 
are shown in Figure~\ref{tt_fig14}.  The temperature distribution is strongly 
peaked, 
although significant outliers exist and range from 0\,K up to 12900\,K.  
In particular, 18\% of the sample, or between 8\% and 38\% if we include the 1$\sigma$ error bars, include measured cloud temperatures that 
are $<\,5000$\,K, which is in the thermally unstable regime \citep{mckee77}.  
\citet{ch01} discussed a survey that is currently obtaining Zeeman splitting 
observations of \ion{H}{1} 21\,cm ISM absorption lines that may be able to provide 
a distribution of temperatures for more distant warm ISM 
environments.  Our turbulent velocity distribution is likewise peaked, and 
the range of values extends from 0\,km\,s$^{-1}$ to 5.6\,km\,s$^{-1}$.  The 
few very high turbulent velocity measurements are likely 
caused by unresolved blends.  Because line 
widths can increase greatly when there are unresolved closely-spaced velocity 
components, 
which are common in the LISM \citep[cf.][]{welty96},
undetected blends will 
increase the widths of all observed ions,  
although only by a small amount for deuterium. 
Therefore, the gas associated with a blended 
absorption feature will be interpreted as having an erroneously high 
turbulent velocity.

Numerous circumstances 
specific to each observation can influence the 
precision of our temperature and turbulent velocity measurements.  For 
example, a single absorption component along the line of sight, high 
resolution and high signal-to-noise observations, and a favorable \ion{H}{1} 
line profile that leads to a straightforward continuum placement for 
\ion{D}{1}, all contribute to a precise measurement of $T$ and $\xi$.  
On the other hand, uncertainties in the derived values of $T$ and $\xi$ are 
generally larger for lines of sight with several velocity components.
Because of the range in precision of our measurements, we weight the mean 
inversely to the measurement error, 
\begin{equation} 
\langle T^{\prime} \rangle = 
\frac{\sum\limits_{i=1}^n(T_i/\sigma_i^2)}{\sum\limits_{i=1}^n(1/\sigma_i^2)},
\label{wmean}
\end{equation}
where $T_i$ is the measured temperature for an absorbing cloud along a line 
of sight, $\sigma_i$ is the associated 1$\sigma$ error, and $n$ is the total 
number of measurements, in this sample $n=50$.  In many cases the upper 
($\sigma_i^+$) and lower ($\sigma_i^-$) bounds of the 1$\sigma$ error do not 
have the same magnitude.  Therefore, when $\sigma_i^+\,\neq\,\sigma_i^-$, we 
use the larger of the two quantities as $\sigma_i$.  The weighted mean 
temperature of 
the observed clouds in the LISM is 
$\langle\,T^{\prime}\,\rangle\,=\,6680$\,K.  Likewise, the mean turbulent 
velocity weighted inversely by the measurement error can be 
calculated from 
Equation~\ref{wmean} by replacing $T$ by $\xi$.  The weighted mean turbulent 
velocity of clouds in the LISM is 
$\langle\,\xi^{\prime}\,\rangle\,=\,2.24$\,km\,s$^{-1}$.

Although the temperature and turbulent velocity distributions shown in 
Figure~\ref{tt_fig14} are significantly peaked about the weighted mean, the 
distributions are not consistent with a constant value of $T$ or $\xi$ for all 
absorbers in the LISM.  This is shown by calculating the $\chi^2_{\nu}$ 
goodness-of-fit measure under the assumption of a constant $T$ and $\xi$ 
model for the LISM.  The minimum value of 
$\chi^2_{\nu}(T_{\rm LISM})\,=\,2.6$ corresponds, as expected, to the 
weighted mean temperature, $\langle\,T^{\prime}\,\rangle\,=\,6680$\,K, and 
likewise, the minimum value of $\chi^2_{\nu}(\xi_{\rm LISM})\,=\,5.6$ 
corresponds to the weighted mean turbulent velocity, 
$\langle\,\xi^{\prime}\,\rangle\,=\,2.24$\,km\,s$^{-1}$.  
However, the large values of 
$\chi^2_{\nu}$ indicate that the observed variations of $T$ and $\xi$ cannot 
be explained by the measurement errors alone, but instead 
the temperature and turbulent velocity 
must vary among collections of gas in the LISM.

In order to quantify the observed variation in temperature and turbulent 
velocity in LISM 
absorbers, we calculate the weighted dispersion about the 
mean value, 
\begin{equation}
\sigma^2_{\langle T^{\prime} \rangle} = 
\frac{\sum\limits_{i=1}^{n}(1/\sigma_i^2)(T_i - \langle T^{\prime} \rangle)^2}
{\sum\limits_{i=1}^{n}(1/\sigma_i^2)} \cdot \left(\frac{n}{n-1}\right),
\label{wvar}
\end{equation}
as described by \citet{bev92}.  Applying Equation~\ref{wvar} to our 
temperature measurements, we find that the dispersion about the weighted mean 
temperature is $\sigma_{\langle\,T^{\prime}\,\rangle}\,=\,1490$\,K.  By 
substituting 
$\xi$ for $T$ in Equation~\ref{wvar},
we find that the dispersion about the 
weighted mean turbulent velocity in LISM absorbers is 
$\sigma_{\langle\,\xi^{\prime}\,\rangle}\,=\,1.03$\,km\,s$^{-1}$.  
Henceforth, any 1$\sigma$ error bars associated with temperature or turbulent 
velocity do not indicate the error in the weighted mean, as this assumes that 
the sample comes from the same parent population or 
equivalently that these 
quantities are constant in the LISM and our sample observations 
would be noisy measurements of that constant 
quantity.  Instead, the 1$\sigma$ errors 
indicate the dispersion about the weighted mean.  In Figure~\ref{tt_fig14} 
the shaded 
distributions indicate subsets of the most precise measurements 
in our sample, those with the smallest measurement errors.  Comparing the 
full distributions to the best measurement subset 
distributions clarifies the 
effect of weighting the mean and 
the dispersion about the mean by the inverse of 
the measurement errors.  
Since no information about the measurement errors is conveyed, 
the complete distribution is implicitly evaluated as if 
all measurements were made with the same precision.  The best measurement 
subset distributions for both $T$ and $\xi$ are 
more strongly peaked than the
complete samples, because 
fewer outliers are included, 
thereby reducing the dispersion.  Because the weighting favors the most 
precise measurements, the mean and dispersion about the mean 
closely correspond to the best measurement subsample distribution.

\subsection{Negative Correlation of Temperature and Turbulent Velocity \label{tt_cor}}

In Figure~\ref{tt_fig15} the temperature and turbulence measurements are 
compared in aggregate for all 50 absorbers.  The subsample of filled symbols 
indicates a selection of the most precise measurements.  The $\pm\,1\sigma$ 
error bars of some of our measurements can span almost the entire parameter 
space, and therefore do not provide much insight into the physical structure 
of the LISM along that particular line of sight.  
We have therefore selected 
a subsample of the most precise measurements, where the absolute error is 
less than the standard deviation of the entire sample, given above.  This 
subsample includes more than a third of the entire sample.  There appears to 
be a moderate negative correlation between the temperature and turbulent 
velocity.  The correlation coefficient ($r$) is --0.47 for the entire sample 
and --0.35 for the precise subsample, with the probability that the 
distribution could be 
drawn from an uncorrelated parent population ($P_c$) 
is 0.010\% and 3.5\%, respectively (assuming no correlation of the errors).  The probability, $P_c$, increases when either $|r|$ or the sample size decreases.  

As is evident from the alignment of the error contours in Figure~1, the 
majority of temperature and turbulent velocity determinations show a 
correlation 
in their measurement errors.  This covariance 
has approximately the 
same alignment as the negative correlation.  In other words, irrespective of 
the presence or absence of a negative correlation, the covariance between the 
errors in $T$ and $\xi$
will tend to smear out the measurements along a negatively sloped 
axis.  To determine the significance of this effect on our sample, we 
calculate a large number of realizations ($\sim\,10^5$) for two 
different scenarios that have the same weighted mean and dispersion in temperature and 
turbulent velocity as our true dataset.  A hypothetical dataset is used, which has the same number of elements as our true sample, but displays no correlation ($r\,=\,0$).  Each realization is a modification of the hypothetical dataset consistent with the errors of our true sample.  The two scenarios correspond to assumption that the errors are: (1) uncorrelated, where the axes of the error ellipses would be parallel to the $T$ and $\xi$ axes, and (2) correlated, as is the case for our true sample, where the error ellipses are not aligned with the principal axes.  The resulting distributions are shown in the bottom of 
Figure~\ref{tt_fig15}.  The first scenario (solid curve) is a distribution of 
samples drawn from uncorrelated errors, which as expected, peaks at $r\,\sim\,0.0$, and falls 
off dramatically at levels of moderate correlation ($|r|\,\sim\,0.3$).  The
second scenario (dashed curve) also draws from our uncorrelated hypothetical sample 
but is varied according to the errors, 
where the errors are drawn from the true covariance matrix, 
taking into account the correlation of measurement errors.  As expected, the distribution 
is slightly skewed to more negative correlations, peaking at $r\,\sim\,-0.19$, due to the smearing of the 
sample along a negatively sloped axis.  
The negatively sloping error ellipses in our sample lead to a shift of any distribution of correlations to slightly more negative values, thereby increasing the probability that the observed sample could be drawn from an uncorrelated parent population.  In this case, the probability increases by about a factor of 10 for the second scenario compared to the first scenario where the errors are uncorrelated (see shaded regions in Figure~\ref{tt_fig15}), but the overall probability still remains relatively small ($<\,1$\,\%).  
Therefore, the covariance between temperature and turbulent velocity is a 
minor influence on the distribution of the sample, and the negative 
correlation between 
$T$ and $\xi$ appears to be significant.

A negative correlation between our temperature and turbulent velocity measurements 
implies that they may not be independent variables and that 
either some physical 
phenomenon, possibly pressure equilibrium, is causing the anticorrelation, or that 
systematic errors in our analysis are leading to the negative correlation.  
Possible sources of systematic errors 
that are unique to each individual spectrum 
include the placement of the stellar continuum and the wavelength 
calibration.  However, we expect these systematic errors to ``randomize'' 
within our sample, because they are different for each spectrum as a result 
of observing a variety of stars, where the location of the absorption is in a 
different part of the stellar continuum, and our use of different instruments 
with a range of spectral resolving powers.  One important source of 
systematic error that 
could be ubiquitous in our sample, is the possible 
presence of unresolved blends.  We attempt to minimize this problem by 
observing the narrowest LISM absorption lines at the highest available 
spectral resolution.  However, one must consider the possible presence of 
several discrete absorbers with very similar projected velocities 
\citep[cf.][]{welty96} and their 
effect on our results when we assume that there is only one absorber along 
the line of sight.  

If unresolved blends are common in our sample, and multiple discrete 
absorbers are interpreted as single absorbers, then our line width 
measurements will be too broad.
Because the unresolved blends will affect all 
resonance lines, 
independent of mass, the overestimate of the line width 
for each ion leads to an overestimate of the turbulent velocity.  Therefore, 
some of the high turbulence measurements are probably due to unresolved 
blends
rather than to intrinsically high turbulent velocities.  If we neglect the 
turbulent velocity measurements greater than 3\,km\,s$^{-1}$, for example, the 
negative correlation between temperature and turbulence is still present.  
The correlation coefficient is then $r\,=\,-0.35$ for 
the full sample and $r\,=\,-0.44$ for our best measurements subsample, and 
the probability ($P_c$) that the distribution results from an uncorrelated 
sample increases to 0.50\% and 2.0\%, respectively.  The weak negative 
correlation may indicate a physical relationship between the temperature and 
turbulence in LISM clouds, perhaps involving pressure equilibrium.

\subsection{Temperature, Turbulent Velocity, and Cloud Mass}

The subtle connections between various cloud properties may help elucidate 
fundamental characteristics of interstellar gas.  In Figure~\ref{ndtxi}, we 
compare the measured temperatures and turbulent velocities with a proxy for 
cloud mass, the \ion{D}{1} column density.  The \ion{D}{1} measurements are 
taken from \citet{red03a}.  As discussed in Section~\ref{calctxi}, a line 
width measurement of a light ion, in particular \ion{D}{1}, is critical 
for an 
accurate determination of cloud temperature and turbulent velocity.  
Therefore, a \ion{D}{1} column density is available for all of the sightlines 
discussed here.  The deuterium-to-hydrogen (D/H) ratio appears to be constant 
within the Local Bubble, D/H$\,=\,(1.56\,\pm\,0.04)\,\times\,10^{-5}$ 
\citep{wood04}.  Therefore, the deuterium column density can be used to 
estimate the hydrogen column density, which is itself a proxy for the mass of 
the cloud along the line of sight, if we assume 
that all these clouds have comparable densities.  

The plots in Figure~\ref{ndtxi} explore whether 
or not the temperature and turbulent 
velocity of the cloud 
vary with the cloud mass. 
Kinetic energy equipartition arguments suggest that, the turbulent motions of smaller clouds 
may be 
greater than those of more massive clouds.  However, no 
correlation is discernible between column density and turbulent velocity, and 
equipartition does not seem to be important on 
the size scales of these clouds.  
We note that a single interstellar cloud may not 
have a unique turbulent velocity.  Along a line of sight through an 
interstellar cloud, it is 
more likely that we are averaging over many turbulent 
environments within a single cloud structure, which 
could mask any equipartition 
signature.  The presence of unresolved velocity structures along the line of
sight could also disrupt the distribution of turbulent velocity measurements.
No correlation is detected between temperature and column density 
either.  Although the full sample seems to show a moderate positive 
correlation, that trend is not mirrored in the best measurement subsample, 
which shows no variation 
in $T$ or $\xi$ with column density.  These comparisons seem to 
indicate that independent of cloud mass or size, the warm partially ionized 
clouds that populate the LISM share similar temperature and turbulent 
properties.

\subsection{Cloud Structure and the Routly-Spitzer Effect}

Implicit in our determination of the temperature and turbulent velocity 
for each component along a line of sight is 
the assumption 
that all the ions used in our analysis are members of the same collection of 
gas.  Because all of the ions used in this paper are expected to be the 
dominant ionization stage, where the ionization fraction is $>50$\%, for physical conditions appropriate for the LISM 
\citep{slavin02}, this assumption is probably fair.  
Although some ions, such as, \ion{C}{2}, \ion{Al}{2}, \ion{Si}{2}, and \ion{Fe}{2}, clearly dominate the ionization distribution with $>95$\% of atoms expected in their ionization stage, other ions, including, \ion{D}{1}, \ion{N}{1}, \ion{O}{1}, and \ion{Mg}{2}, have less extreme ionization fractions ranging from 55\%-85\%.  \citet{wood02} used LISM column density measurements along the line of sight toward Capella to make a direct comparison between observed ionization fractions of carbon, nitrogen, and silicon, and model predictions by \citet{slavin02}.  
If differential ionization is acting such that various ions trace distinct interstellar medium, particularly since the degree of ionization fraction dominance spans a significant range for the observed ions, 
the measured line widths may not fit a single temperature and turbulent velocity.
We can 
test the assumption that the observed ions reside in the same collection of gas by looking for systematic deviations 
in the Doppler width of each ion compared with the trend 
for other ions in the same component
that would arise if individual ions were sampling different environments.  
For example, along distant lines of sight ($\sim\,1\,$kpc), \citet{routly52} 
noted that the velocity dispersion of the absorbers increased for clouds 
which had low \ion{Na}{1} to \ion{Ca}{2} ratios.  This has been confirmed by 
more recent observations \citep{siluk74,vallerga93}, and 
is attributed to large 
scale structures, such as old supernova shells, that can account for both the 
large velocity dispersion and the 
change in abundance ratio.  It is possible 
that something analogous to the Routly-Spitzer effect, whether 
due to varying 
depletion or ionization structure within an individual cloud, is acting on a 
much smaller scale, 
thereby separating the observed ions into slightly different 
environments within the cloud.  We investigate this issue briefly here, while a detailed discussion of the ionization structure of the LISM will be provided in a subsequent paper.

In Figure~\ref{disper}, we 
test for any systematic variations by
showing the weighted mean deviation for each ion 
from the best fit temperature and turbulent velocity solution for each cloud. 
The results are plotted 
against atomic mass as well as 
depletion for a typical diffuse cloud
taken from the cool diffuse cloud along the $\zeta$~Oph sightline \citep{savage96}.  Although the magnitudes of the depletions of the two diffuse clouds along the line of sight toward $\zeta$~Oph are slightly offset from each other, the general trend in Figure~\ref{disper} is the same regardless of which cloud is used as the depletion standard.  
The absorption features of 
both 
carbon and oxygen are typically saturated in our sample, which leads to their large 
dispersion about the weighted mean.  Both deuterium and iron show very little deviation from the best fit value because they are at opposing ends of the range of atomic masses, and therefore, are the most influential in the best fit determination of the temperature and turbulent velocity.
There is no indication of a systematic 
trend for different ions based on their depletion.  In fact, the weighted 
mean of $\Delta$b from the best fit value is never greater than 
0.1\,km\,s$^{-1}$, which supports the assumption that these particular ions 
reside in the same collection of gas and 
provides accurate determinations of 
temperature and turbulent velocity.

\section{Estimate of the Thermal and Turbulent Pressures in the LISM 
\label{secpres}}

From the distribution of the observed 
temperatures and turbulent velocities
in the warm clouds in the LISM, we can estimate the mean thermal and 
turbulent pressures.  The thermal pressure is defined as,
\begin{equation}
P_{\rm T} = nkT,
\end{equation}
where $n$ is the number density of the particles in the gas, $k$ is 
Boltzmann's constant, and $T$ is the 
gas temperature.  Likewise, the 
turbulent pressure is defined as,
\begin{equation}
P_{\xi} = \frac{1}{2}\rho\xi^2,
\end{equation}
where $\rho$ is the mass density of the 
gas particles, and $\xi$ is 
the turbulent velocity of the gas.  In order to calculate the 
gas density in the LISM, we assume cosmic abundances, or 
$n({\rm H})/n({\rm He})\,\sim\,10$, and that the number of electrons 
equals the number of protons.  The electron number density is assumed to be 
$n_{\rm e}\,=\,0.11^{+0.12}_{-0.06}$\,cm$^{-3}$, as measured for the Capella 
line of sight through the Local Interstellar Cloud (LIC) using the 
\ion{C}{2}$^*$ fine structure line \citep{wood97}, and using a nearby white 
dwarf \citep{holberg99}.  The hydrogen number density for the LISM is assumed 
to be $n({\rm HI})\,=\,0.1\,$cm$^{-3}$ \citep{red00}.  The number density 
($n$) is calculated from $n\,=\,\sum\,n_i$, where $i$ 
signifies the 
particles included in this computation: \ion{H}{1} ions, electrons, protons, 
and helium ions.  Similarly, the density ($\rho$) is calculated from 
$\rho\,=\,\sum\,n_i\,m_i$.  

Given the mean temperature of the LISM, $T\,=\,6680$\,K and the measured dispersion of 1490\,K, the mean 
thermal pressure is calculated to be 
$P_{\rm T}/k\,=\,2280$\,K\,cm$^{-3}$ with a dispersion about the mean of 520\,K\,cm$^{-3}$, which is 
consistent with the value $P/k= 1700$ to 2300 cm$^{-3}$~K estimated by 
\citet{vall96} from {\it Extreme Ultraviolet Explorer} ({\it EUVE}) spectra.  
The mean turbulent pressure and standard deviation, for a mean turbulent velocity of 
2.24\,km\,s$^{-1}$ and dispersion of 1.03\,km\,s$^{-1}$, is 
$P_{\xi}/k\,=\,89\pm82$\,K\,cm$^{-3}$.  The thermal pressure thus 
dominates the turbulent pressure, $P_{\rm T}/P_{\xi}\,\sim\,26$.  Therefore, 
the small-scale motions in the warm clouds in the LISM, are dominated by 
thermal motions, as opposed to small-scale turbulence.  
The pressure estimates given above ignore the measurement error of $n_{\rm e}$ in order to show the magnitude of the dispersion about the weighted mean pressures.  If instead, we ignore the dispersion about the mean temperature and turbulent velocity, and include the measurement error associated with $n_{\rm e}$, the measured thermal pressure is $P_{\rm T}/k\,=\,2280^{+1770}_{-880}$\,K\,cm$^{-3}$, and the turbulent pressure is $P_{\xi}/k\,=\,89^{+36}_{-18}$\,K\,cm$^{-3}$.
Different estimates 
of the hydrogen number density will not change this ratio significantly 
because the number density essentially cancels out of the ratio.

The mean total pressure of the warm partially ionized clouds that populate 
the LISM is $P_{\rm tot}/k 
=\,P_{\rm T}/k\,+\,P_{\xi}/k\,=\,2370$\,K\,cm$^{-3}$ with a dispersion about he mean of 530\,K\,cm$^{-3}$.  These 
clouds are surrounded by a larger structure, the Local Bubble, which 
presumably is characterized by hot ($T\,\sim\,10^6$\,K) low density 
($n_{\rm e}\,\sim\,0.005$\,cm$^{-3}$) gas, which corresponds to a thermal 
pressure of $P_{\rm T}/k\,\sim\,11000$\,K\,cm$^{-3}$ \citep{sanders77}.  The 
pressure imbalance between the warm and hot components of the LISM has been a 
persistent problem for understanding the structure of our local interstellar 
environment \citep{jenkins02}.  
The present survey demonstrates that this 
discrepancy is common to all 
known local interstellar clouds within 100\,parsecs, 
and that turbulent pressure is too small to make up the difference between 
the thermal pressure differences of the warm and hot LISM components.  Recent 
results have stimulated a reexamination of the soft X-ray emission 
measurements of the Local Bubble, and hence the thermal pressure calculation 
of the hot medium.  It turns out that 25-50\% of the detected soft X-rays, 
initially wholly attributed to the hot gas in the Local Bubble, are the 
result of charge exchange between solar wind ions and interplanetary 
neutrals \citep{cox98,cravens00,robertson03}.  Therefore, the thermal 
pressure of the hot component of the LISM may be significantly lower than the 
initial estimates, and closer 
to agreement with the warm clouds of the LISM.

\section{Turbulent Mach Number}

We calculate the turbulent Mach number ($M_{\xi}$) by taking the ratio of the 
observed turbulent velocity 
to the sound 
speed of the gas along the line of sight,
\begin{equation}
M_{\xi} = \frac{\xi}{\left({\frac{n k T}{\rho}}\right)^{1/2}},
\label{macheq}
\end{equation}
where the calculation of the number density ($n$) and mass density ($\rho$) 
are described in Section~\ref{secpres}.  The results of this calculation are 
shown as a function of cloud temperature in Figure~\ref{tt_f18}.  
The gradual decrease in turbulent Mach number as a function of 
increasing cloud 
temperature that is seen in Figure~\ref{tt_f18}
is a consequence of the increase in sound speed with temperature, as well as the anticorrelation of turbulent velocity with temperature, discussed in Section~\ref{tt_cor}.  
Again, the best measurement 
subsample is displayed as red symbols, almost all of which are clustered at a 
turbulent Mach number of $\sim\,0.2$.  The weighted mean of the turbulent 
Mach number ($\langle\,M^{\prime}_{\xi}\,\rangle$) is calculated by replacing 
$T$ by $M_{\xi}$ in Equation~\ref{wmean}, where 
$\langle\,M^{\prime}_{\xi}\,\rangle\,=\,0.19$.  Similarly, the weighted 
dispersion about the mean value 
($\sigma_{\langle\,M^{\prime}_{\xi}\,\rangle}$) is derived from 
Equation~\ref{wvar}, where 
$\sigma_{\langle\,M^{\prime}_{\xi}\,\rangle}\,=\,0.11$.  The same data are also 
displayed in histogram form in Figure~\ref{tt_f18_5}.  Clearly, the 
macroscopic nonthermal motions of warm clouds in the LISM are significantly 
subsonic, in contrast to the highly supersonic turbulent motions in cold 
neutral clouds \citep{heiles03}, and in agreement with optical observations of cool clouds \citep{welty96,dunkin99}.

\section{Spatial Distribution of Temperature and Turbulent Velocity}

In Figures~\ref{tt_f19} and \ref{tt_f20}, the observed temperature and 
turbulent velocity, respectively, for each line of sight are shown in 
Galactic coordinates.  The size of a symbol is inversely proportional to the 
target's distance, and the color of a symbol indicates the observed quantity, 
as shown by the scale at the bottom of each plot.  Sightlines that contain 
more than one absorber, or sightlines to 
multiple targets, such 
as binary systems, include measurements of all clouds by separating the 
shadings with a dotted line.  In this way, each sightline will have between 
one and three different shadings to signify up to three absorbers along the 
line of sight.

In a number of cases, both the temperature and turbulent velocity seem to be 
spatially correlated.  For example, the closest pair of sightlines, 
$\alpha$~Cen~A and $\alpha$~Cen~B, have almost identical temperature and 
turbulent velocity measurements.  The same is true for another close pair, 
Capella and the more distant white dwarf G191-B2B, which are only separated 
by 7$^{\circ}$.  The $+20$\,km\,s$^{-1}$ component is common in the spectra 
of both stars, and the individual temperature and turbulent velocity 
measurements are practically identical.  On the other hand, a closely grouped 
trio of stars in the direction of the North Galactic Pole, share only 
moderately similar cloud characteristics.  With angular separations of 
5$^{\circ}$-8$^{\circ}$, the observed spectra of 31~Com, HZ~43, and GD~153, 
each show one absorption component 
with roughly the same velocity.  The 
derived temperature measurements are similar, ranging from 7000-8200\,K, but 
the turbulent velocity measurements, while all less than the LISM mean, 
range from no detected turbulent velocity to 
1.7\,km\,s$^{-1}$
but all are formally consistent with $\xi = 0.0$\,km\,s$^{-1}$.
The temperatures and turbulent velocities for the two closely spaced 
velocity components in the lines of sight to Procyon and 61~Cyg~A are identical because the line widths were assumed to be the same due to the blending of the absorption components.

In order to quantify the magnitude of the spatial correlations in our 
measurements, we have compared the 
absolute values of the temperature and turbulent velocity 
differences for all unique pairings of the stars listed in 
Table~\ref{tt_table1}.  The results are plotted in Figure~\ref{tt_f20_5} as a 
function of the angular separation of the pair of sightlines, in 10$^{\circ}$ 
bins.  We also applied this procedure to the 
absolute values of the observed projected velocity 
difference ($\Delta v$) of 
the absorbing cloud and the total pressure, 
which is a function of both 
temperature and turbulent velocity.  

It has been known for some time that the projected velocity of absorbing 
material in the LISM shows a strong spatial correlation \citep{crutcher82}.  
In fact, single three dimensional velocity vectors have been used to describe 
the bulk flow of the closest interstellar clouds, and is one of the 
predominant reasons 
that LISM absorbers are thought of as roughly homogeneous 
cloud-like entities, rather than chaotic and filamentary objects 
\citep{lall92,lall95}.  The spatial correlation in velocity is 
clearly seen  in the 
top panel of Figure~\ref{tt_f20_5}, where both the absolute velocity 
difference, and the dispersion about the weighted mean, are significantly 
less for sightlines with angular separations $<$\,20$^{\circ}$.  The dotted 
lines indicate the weighted mean of measurements from 
0$^{\circ}$-20$^{\circ}$, and then from 20$^{\circ}$-60$^{\circ}$.  The 
absolute value of the weighted mean velocity 
separation for close pairs is 21\% that of more 
distant pairs.  The same trend seems to be true for temperature, where the 
absolute value of the weighted mean temperature 
difference of pairs with separations $<$\,20$^{\circ}$ is 
57\% of the weighted mean of more distant pairs.  For turbulent velocity 
measurements the difference between the closest and 
furthest pairings is only 
marginal, just 81\%.  The total pressure of the cloud shows a significant 
spatial correlation, where the weighted mean of pairings $<$\,20$^{\circ}$ 
apart is 50\% of the weighted mean of pairings with greater angular 
separations.  The spatial correlation of total pressure may be related to the 
negative correlation between temperature and turbulent velocity discussed in 
Section~\ref{tt_cor}.  Although 
temperature and turbulent velocity may 
each vary over short angular distances, pressure equilibrium may be acting to 
minimize the total pressure variation over small angular distances.

Temperature 
should be useful as a unique cloud characteristic, because 
like velocity the 
total measurement range is large but 
over small angular separations the mean temperature range is relatively
small.  However, the correlation of temperature with small angular separations is not as tight as for velocity.
The turbulent 
velocity measurement may have less to do with the gross characteristics of 
a specific interstellar cloud, but could vary greatly inside a cloud from stratification of the particular physical structure of the gas due to 
external forces such as cloud-cloud interactions.

\section{Conclusions}

We present 
here the first systematic survey of temperature and turbulent velocity 
in the warm partially ionized clouds of the LISM.  The collections of 
observations presented by \citet{red02,red03a}, and researchers referenced 
therein, provide an opportunity to make fundamental physical measurements of 
the LISM, including 50 independent temperature and turbulent velocity 
measurements of the gas 
along 29 lines of sight in the LISM within 100 pc.  
The results of this work can be 
summarized as follows:
\begin{enumerate}
\item We find that in order to make an accurate and precise measurement of 
the gas temperature and turbulent velocity, the line widths of deuterium, and 
an ion at least as heavy as magnesium are required.  Without deuterium there 
is no serious constraint on the temperature, and without a relatively heavy 
ion, there is no serious constraint on the turbulent velocity.  
High resolution observations are absolutely essential, because except for 
deuterium, the absorption lines are not resolved at moderate resolution, and 
systematic errors can dominate the measurements, 
leading to erroneous results.
\item The distributions of temperature and turbulent velocity of LISM gas 
located are both significantly peaked.  The weighted mean and 
standard deviation for the temperature is $6680\,\pm\,1490$\,K.  The weighted 
mean and standard deviation for the turbulent velocity distribution is 
2.24$\,\pm\,1.03$\,km\,s$^{-1}$.  The high turbulent velocity measurements 
are most likely due to unresolved blends.  The standard deviation indicates 
the range of variation in the LISM, rather than the measurement error of the 
mean value.  We find a 
significant number (18\%, or between 8\% and 38\% including the 1$\sigma$ errors) of sightlines with temperatures 
$<\,5000$\,K, below the stable WNM temperature regime defined by 
\citet{mckee77}.  
\item A moderately negative correlation is detected between temperature and 
turbulent velocity.  The correlation coefficient ($r$) is --0.47 for the 
entire sample and --0.35 for the precise subsample, with the probability that 
the distribution could be from an uncorrelated parent population ($P_c$) of 
0.010\% and 3.5\%, respectively.   The covariance of the errors of the temperatures and turbulent velocities do not dominate the determination of this negative correlation.
Pressure equilibrium 
among the warm clouds may be the source of this anticorrelation.  
\item The small-scale thermal motions dominate the turbulent motions.  For 
the mean temperature and turbulent velocity, we calculate a mean thermal pressure 
of $P_{\rm T}/k\,=$ 2280\,K\,cm$^{-3}$ and a dispersion of 520\,K\,cm$^{-3}$, which is consistent 
with the value $P/k= 1700$ to 2300\,cm$^{-3}$~K estimated by \citet{vall96} 
from {\it EUVE} spectra.  The 
mean thermal pressure dominates the mean turbulent pressure, 
$P_{\xi}/k\,=\,89$\,K\,cm$^{-3}$ with a dispersion about the mean of 82\,K\,cm$^{-3}$, by more than an 
order of magnitude, $P_{\rm T}/P_{\xi}\,\sim\,26$.  Turbulent pressure in 
LISM clouds cannot make up the difference in the apparent pressure imbalance 
between warm LISM clouds and the surrounding hot gas of the Local Bubble. 
\item We find that the turbulent motions of the warm partially ionized clouds 
in the LISM are significantly subsonic.  The weighted mean and standard 
deviation of the turbulent Mach number for the entire sample is 
0.19\,$\pm$\,0.11.  Although some turbulent Mach numbers approach the sound 
speed, due to relatively low temperatures of the clouds, these 
temperatures are not accurately known.  The best measurement subsample 
clusters almost exclusively around turbulent Mach numbers of $\sim\,0.2$.
\item We have calculated the spatial distribution of the temperature and 
turbulent velocity for the 50 
velocity components in the LISM.  These individual 
measurements can now be combined with additional physical properties of LISM 
clouds, such as projected velocity and depletions, to determine the 
three-dimensional morphological structure of gas within 100\,pc. 
This work will be presented in subsequent papers.

\end{enumerate}

\acknowledgments
Support for program AR-09525.01A is provided by NASA through a grant to the 
University of Colorado at Boulder from the Space Telescope Science Institute, 
which is operated by the Association of Universities for Research in 
Astronomy, Inc., under NASA contract NAS AR-09525.01A.  This work is also 
supported by a NASA GSRP student fellowship grant NGT 5-50242.  This research has made use of the SIMBAD database,
operated at CDS, Strasbourg, France.  This research has made use of NASA's Astrophysics Data System.  We would 
like to thank Tom Ayres, Mike Shull, Brian Wood, Edward Robinson, and Steve Federman for 
their helpful comments and suggestions.  We are also indebted to the careful 
reading and excellent comments from the referee.

\clearpage
\begin{figure}
\epsscale{.4}
\plotone{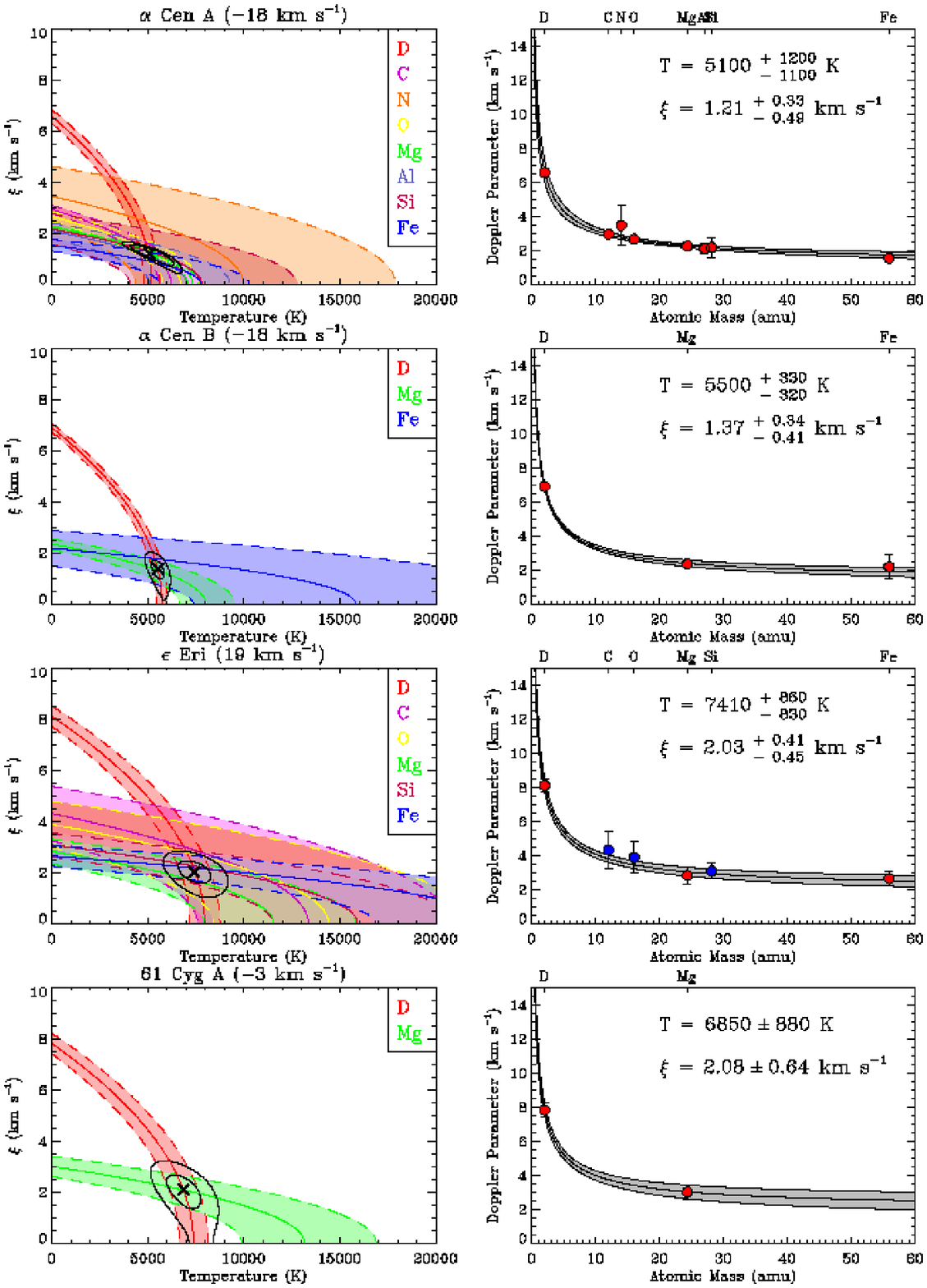}
\figurenum{1a}
\caption{Determination of the temperature ($T$) and turbulent velocity 
($\xi$) for each LISM absorption component observed within 100\,pc.  These 
measurements require an observation of a light ion such as \ion{D}{1}, and at 
least one other ion at least as heavy as \ion{Mg}{2}.  Each LISM component is 
labeled in the plot on the left with the sightline designated by the 
background star, and the velocity of the LISM absorption.  The plot on the 
left shows the best fit Doppler parameter (solid lines) and $\pm\,1\sigma$ 
error bars (dashed lines) for each ion observed, colored coded and labeled 
toward the right side of the plot.  Each curve is determined by the measured 
Doppler parameter and mass of the observed ion, as described in 
Equation~\ref{tt_eq1}.  The ``$\times$'' symbol indicates the best fit 
temperature and turbulent velocity given the widths of all observed ions, 
while the surrounding black contours indicate the $\pm\,1\sigma$ and 
$\pm\,2\sigma$ error bars.  The plot on the right shows the measured Doppler 
parameters as a function of the atomic mass of the observed ion, which are 
specifically indicated in the top axis.  Red symbols indicate that the LISM 
absorption line was measured from a high resolution observation and used to 
make the fit, and blue symbols indicate that the measurements were from 
moderate resolution observations and are used as a consistency check.  The 
best fit temperature and turbulent velocity is given in the 
right plot, and the 
best fit curve is displayed by the solid line, and the shading includes the 
$\pm\,1\sigma$ range of possible outcomes given the derived $T$ and $\xi$ 
parameters and their respective errors. \label{tt_fig1}}
\end{figure}

\clearpage
\begin{figure}
\epsscale{.9}
\figurenum{1b}
\plotone{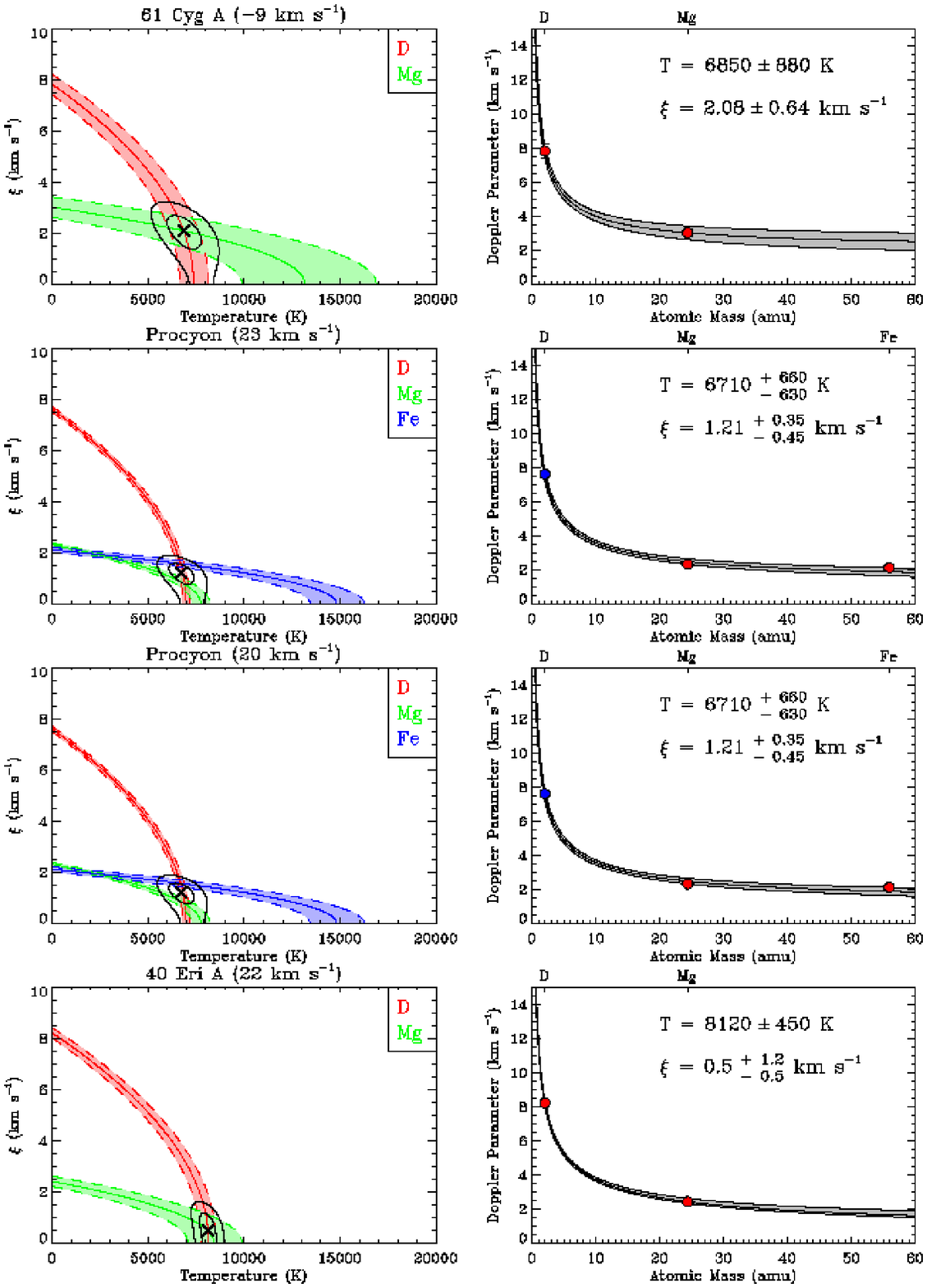}
\caption{Continuation of Figure~\ref{tt_fig1}. \label{tt_fig2}}
\end{figure}

\clearpage
\begin{figure}
\epsscale{.9}
\figurenum{1c}
\plotone{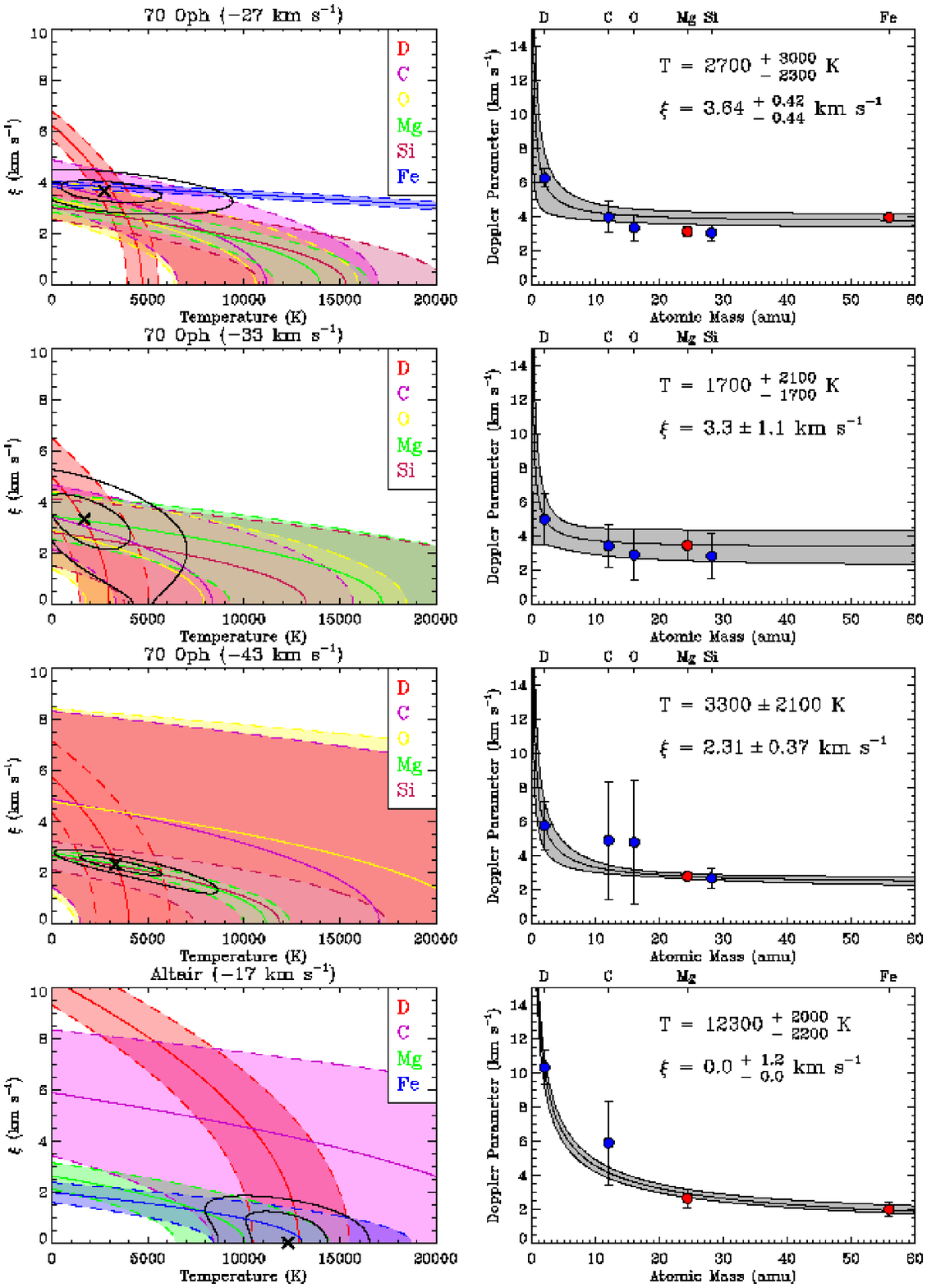}
\caption{Continuation of Figure~\ref{tt_fig1}. \label{tt_fig3}}
\end{figure}

\clearpage
\begin{figure}
\epsscale{.9}
\figurenum{1d}
\plotone{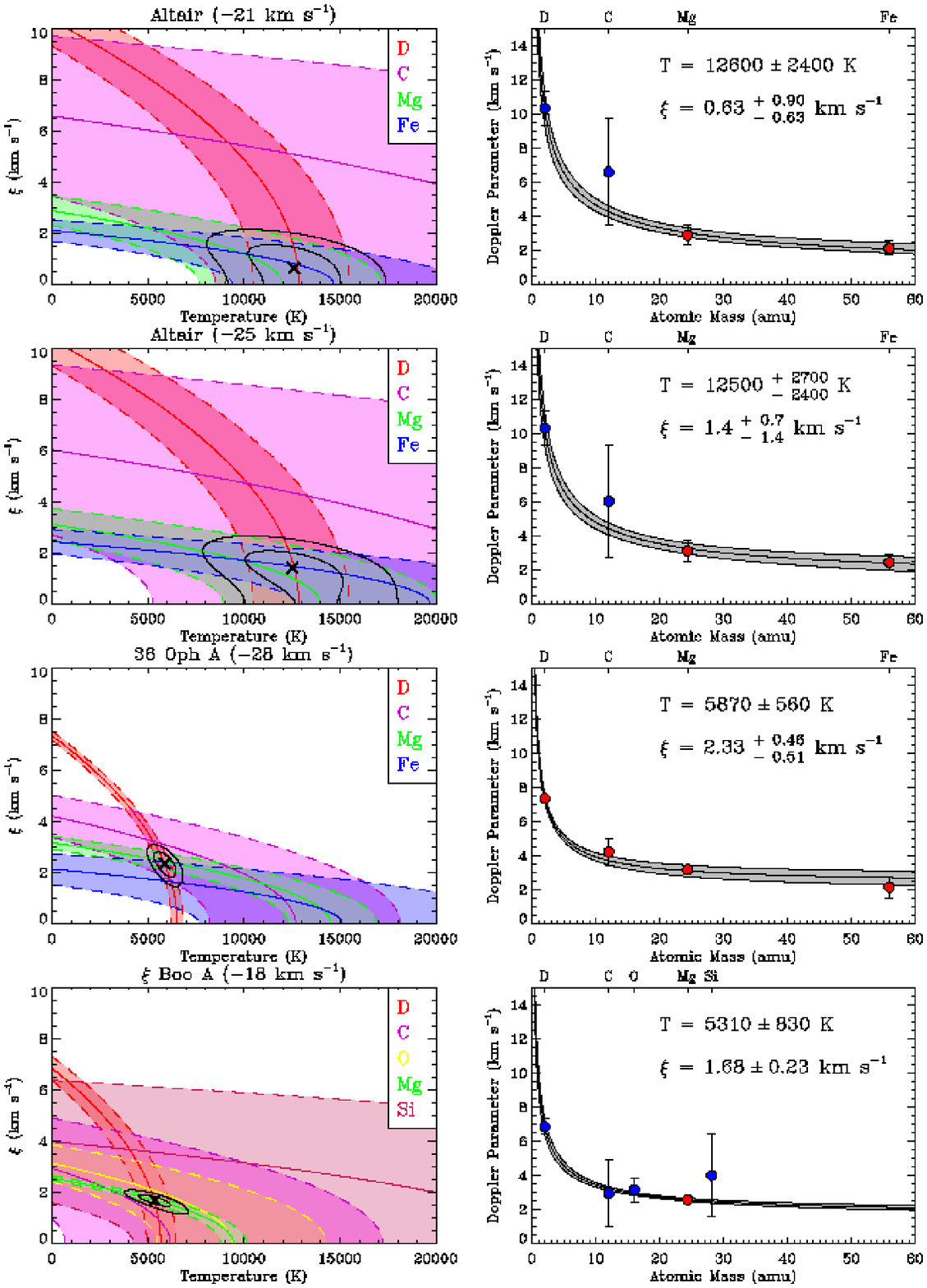}
\caption{Continuation of Figure~\ref{tt_fig1}. \label{tt_fig4}}
\end{figure}

\clearpage
\begin{figure}
\figurenum{1e}
\epsscale{.9}
\plotone{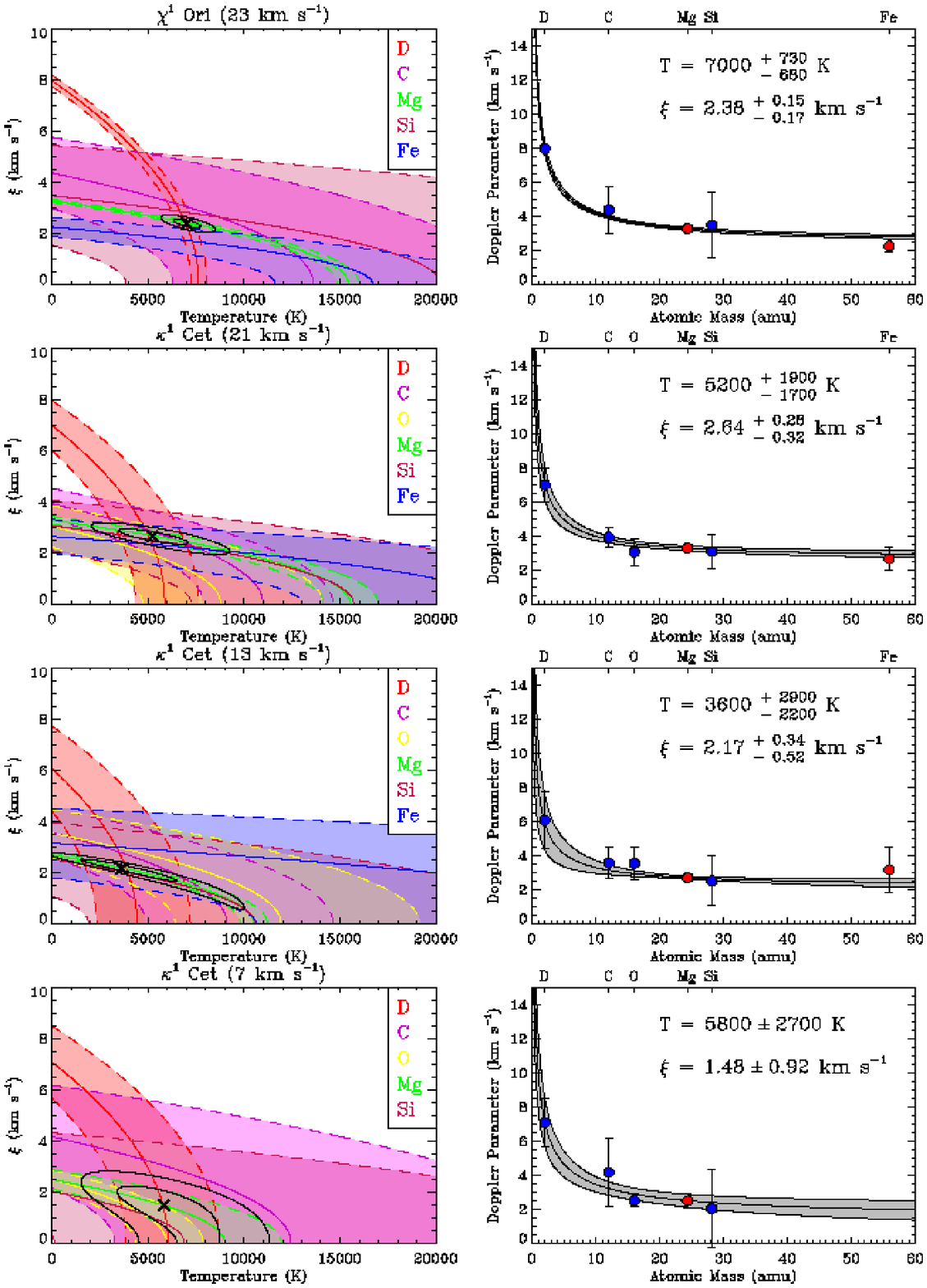}
\caption{Continuation of Figure~\ref{tt_fig1}. \label{tt_fig5}}
\end{figure}

\clearpage
\begin{figure}
\figurenum{1f}
\epsscale{.9}
\plotone{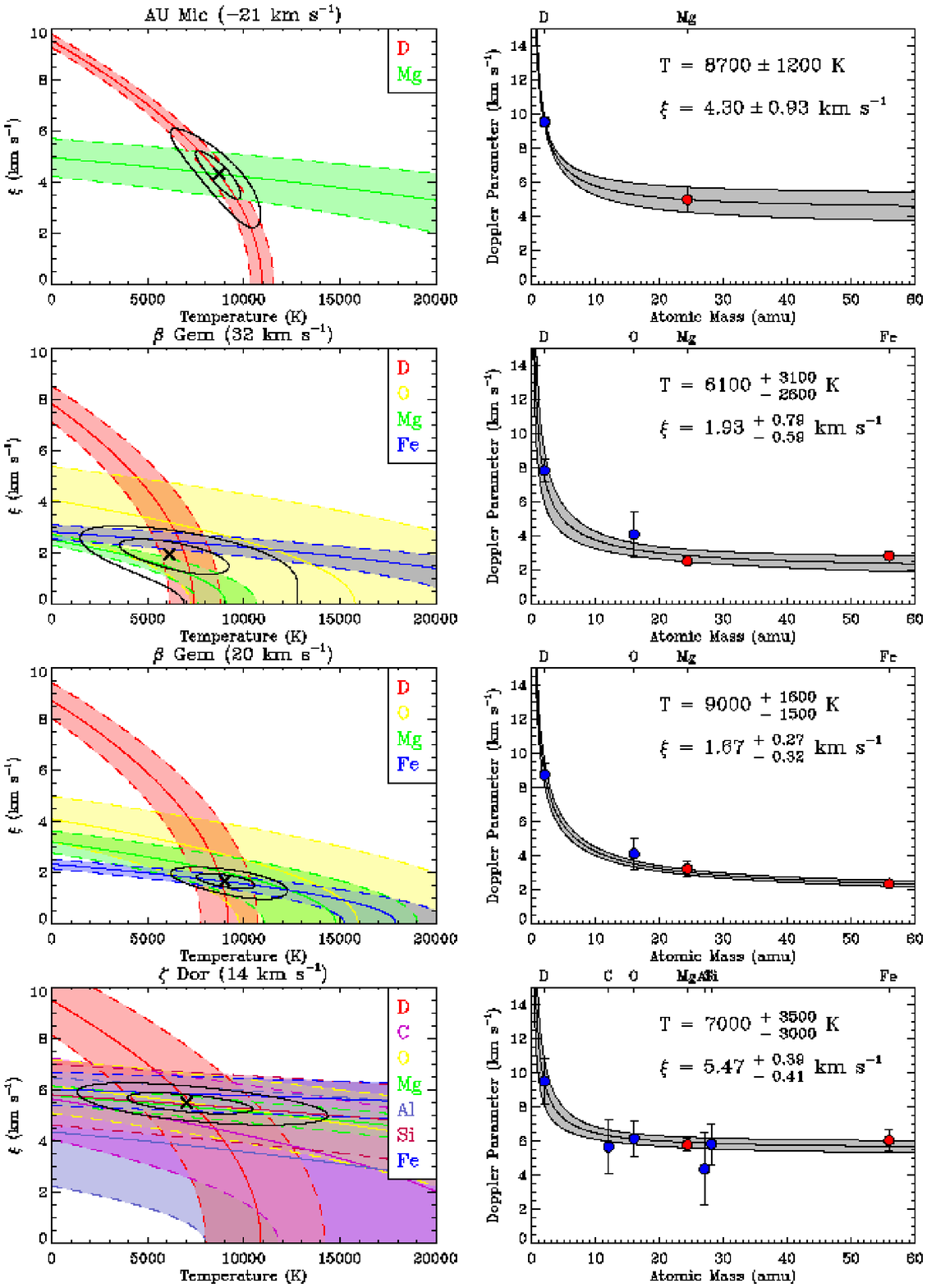}
\caption{Continuation of Figure~\ref{tt_fig1}. \label{tt_fig6}}
\end{figure}

\clearpage
\begin{figure}
\figurenum{1g}
\epsscale{.9}
\plotone{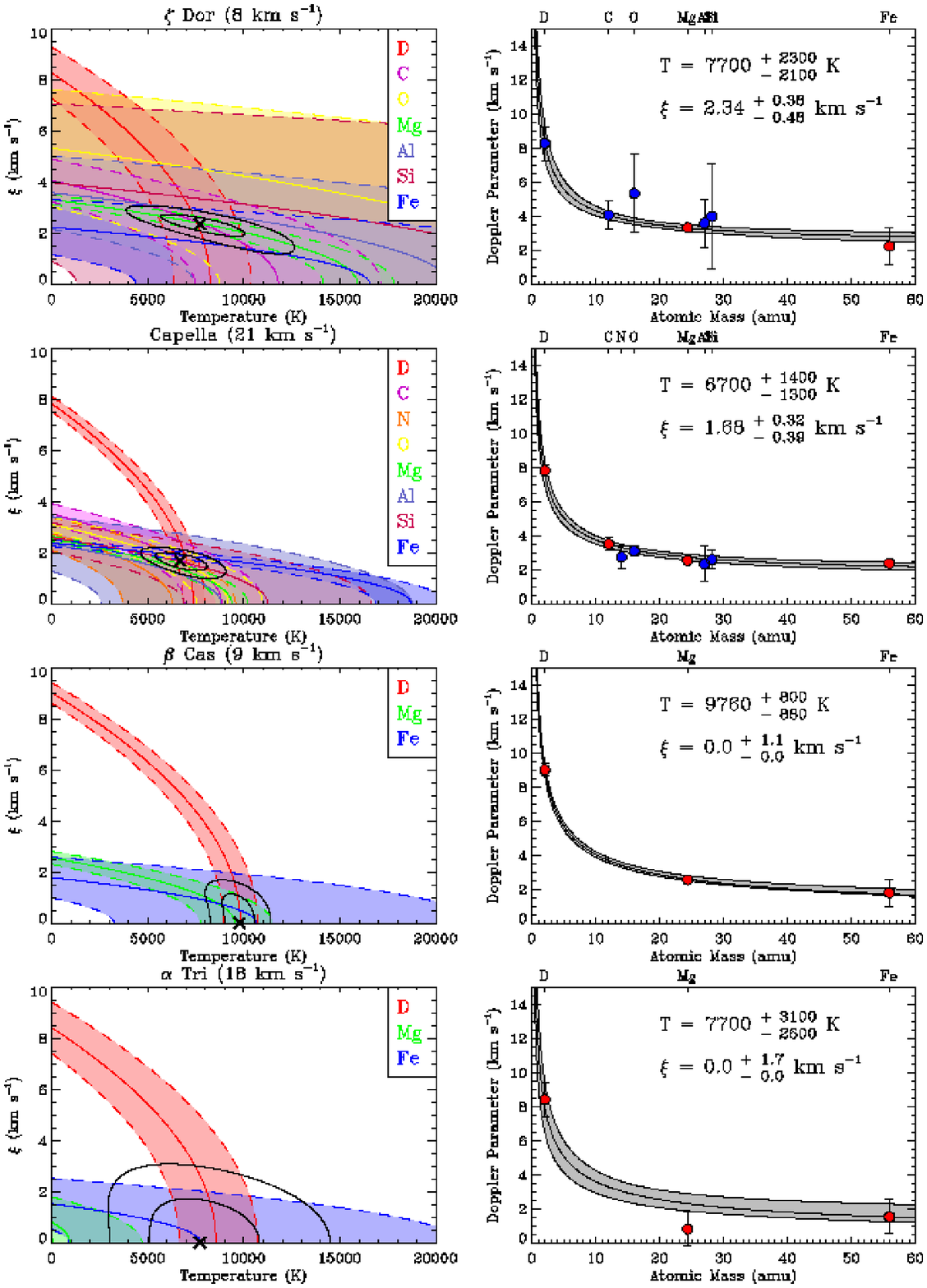}
\caption{Continuation of Figure~\ref{tt_fig1}. \label{tt_fig7}}
\end{figure}

\clearpage
\begin{figure}
\figurenum{1h}
\epsscale{.9}
\plotone{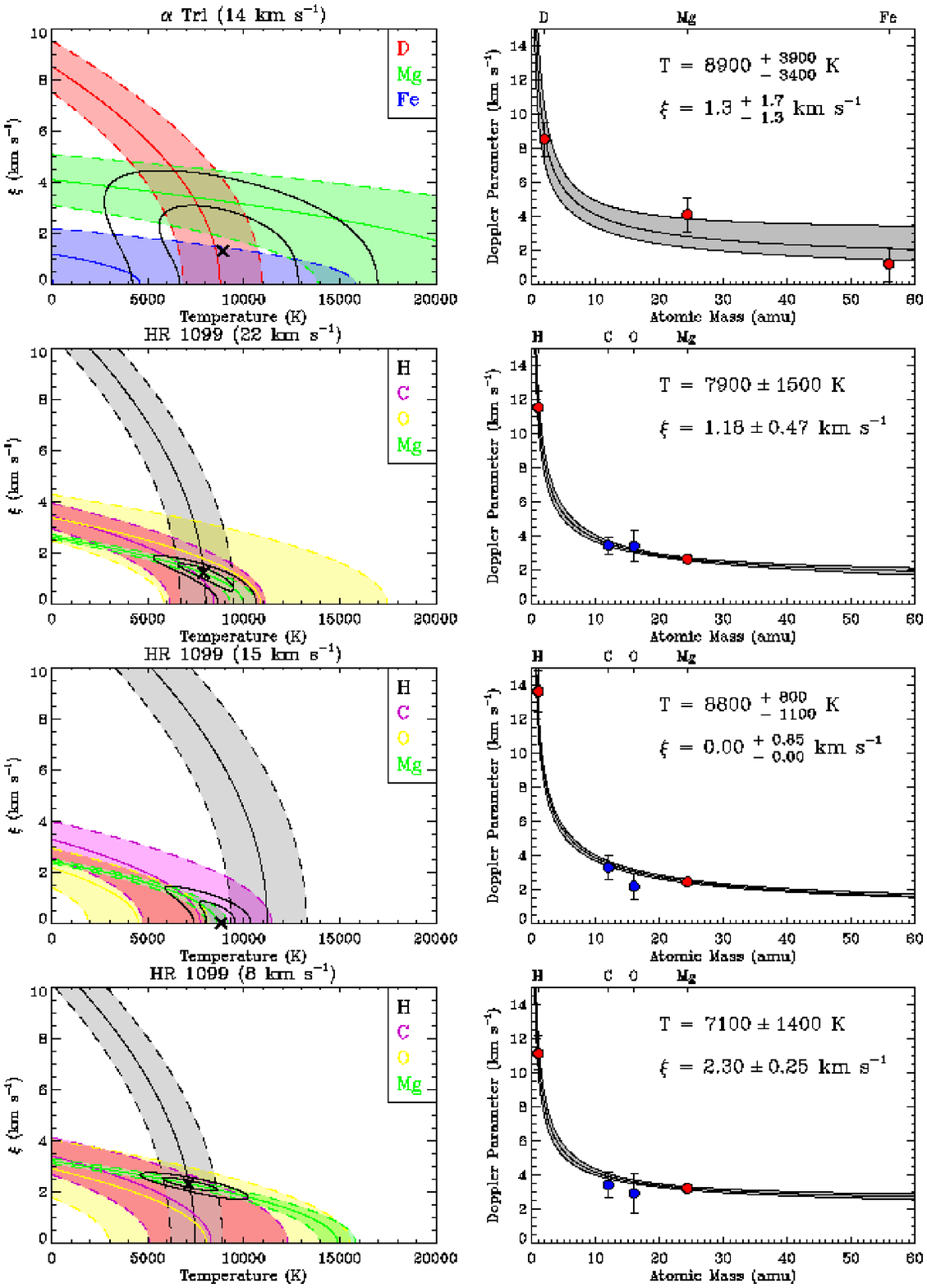}
\caption{Continuation of Figure~\ref{tt_fig1}. \label{tt_fig8}}
\end{figure}

\clearpage
\begin{figure}
\figurenum{1i}
\epsscale{.9}
\plotone{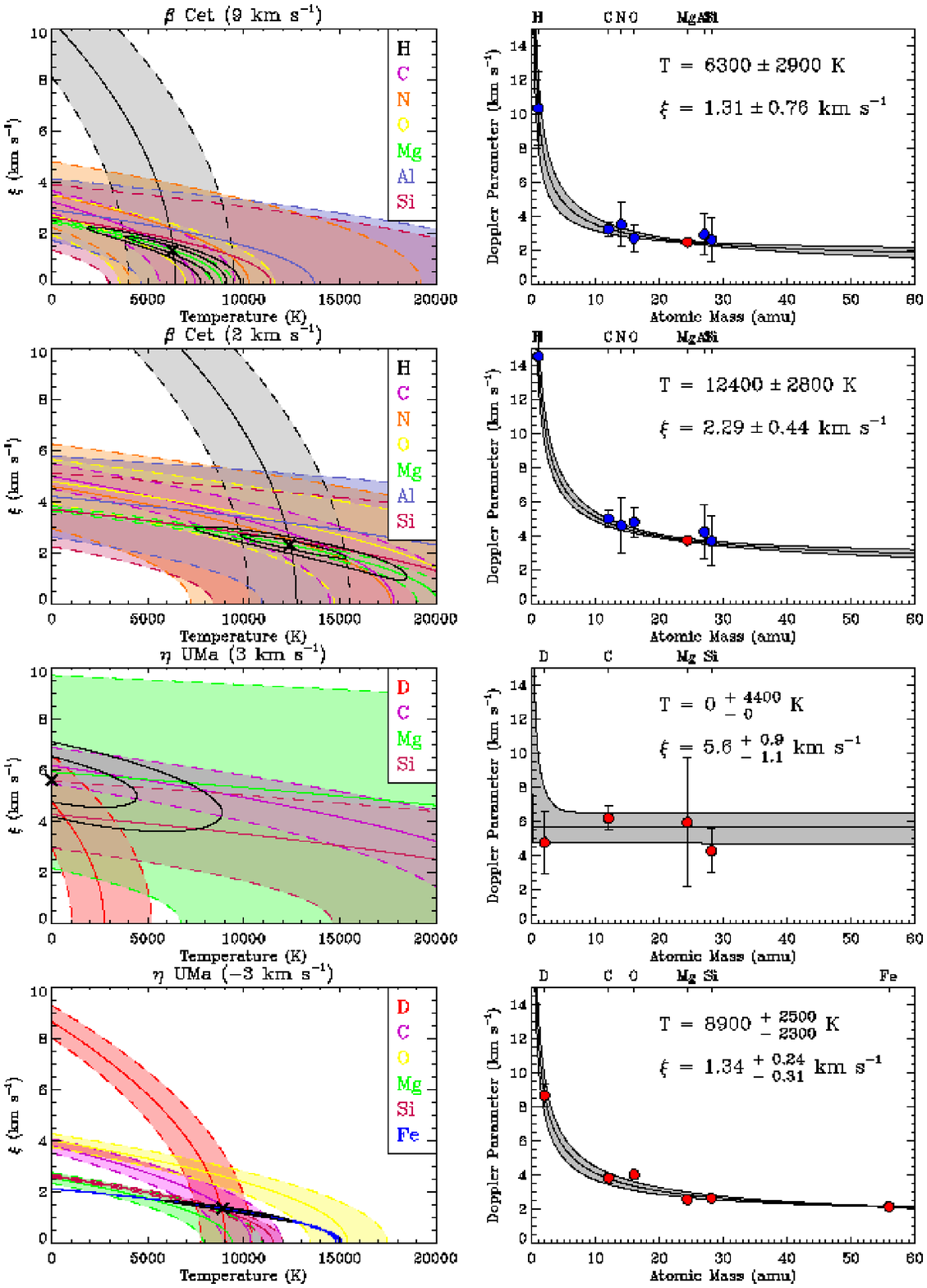}
\caption{Continuation of Figure~\ref{tt_fig1}. \label{tt_fig9}}
\end{figure}

\clearpage
\begin{figure}
\epsscale{.9}
\figurenum{1j}
\plotone{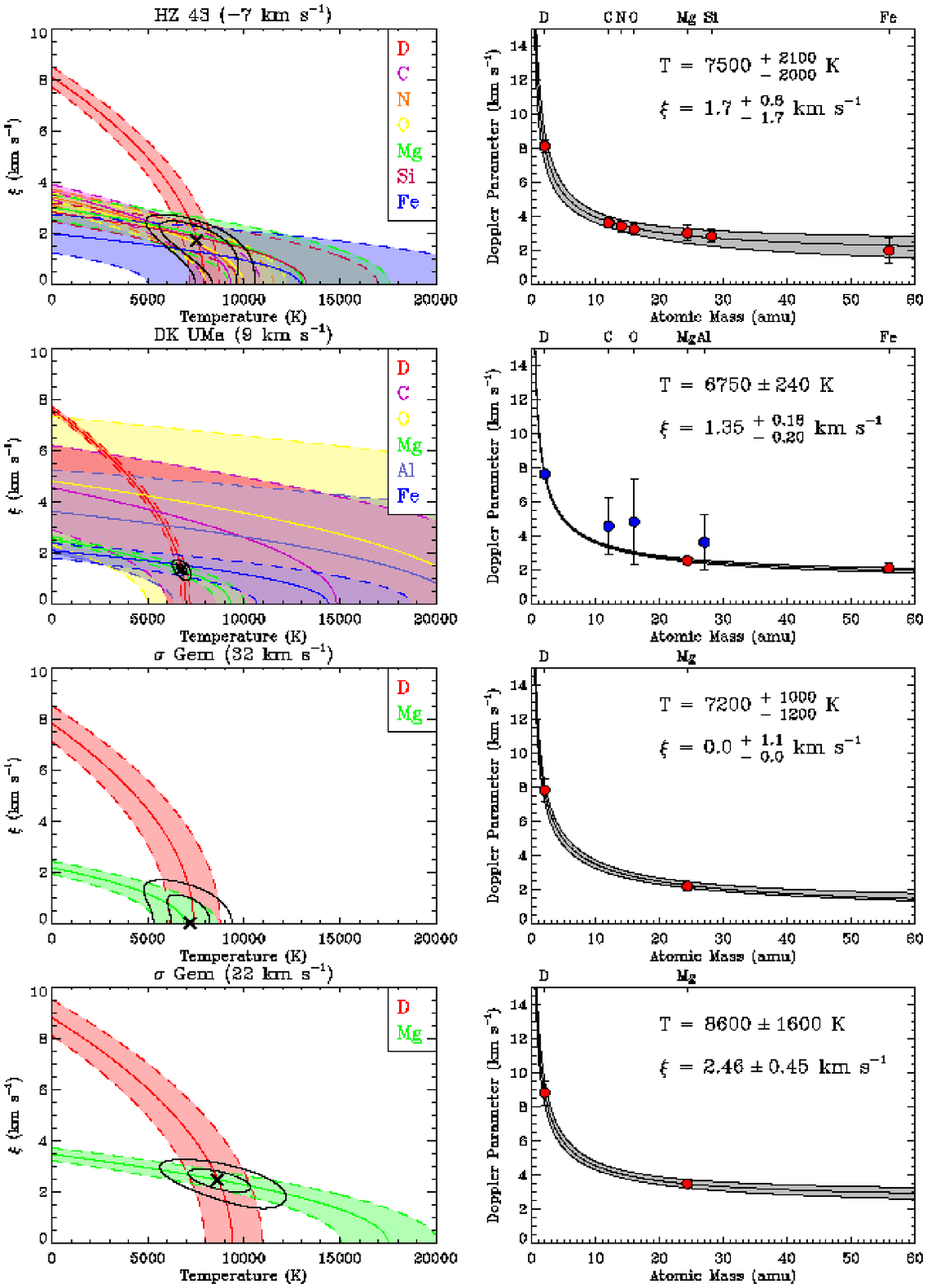}
\caption{Continuation of Figure~\ref{tt_fig1}. \label{tt_fig10}}
\end{figure}

\clearpage
\begin{figure}
\epsscale{.9}
\figurenum{1k}
\plotone{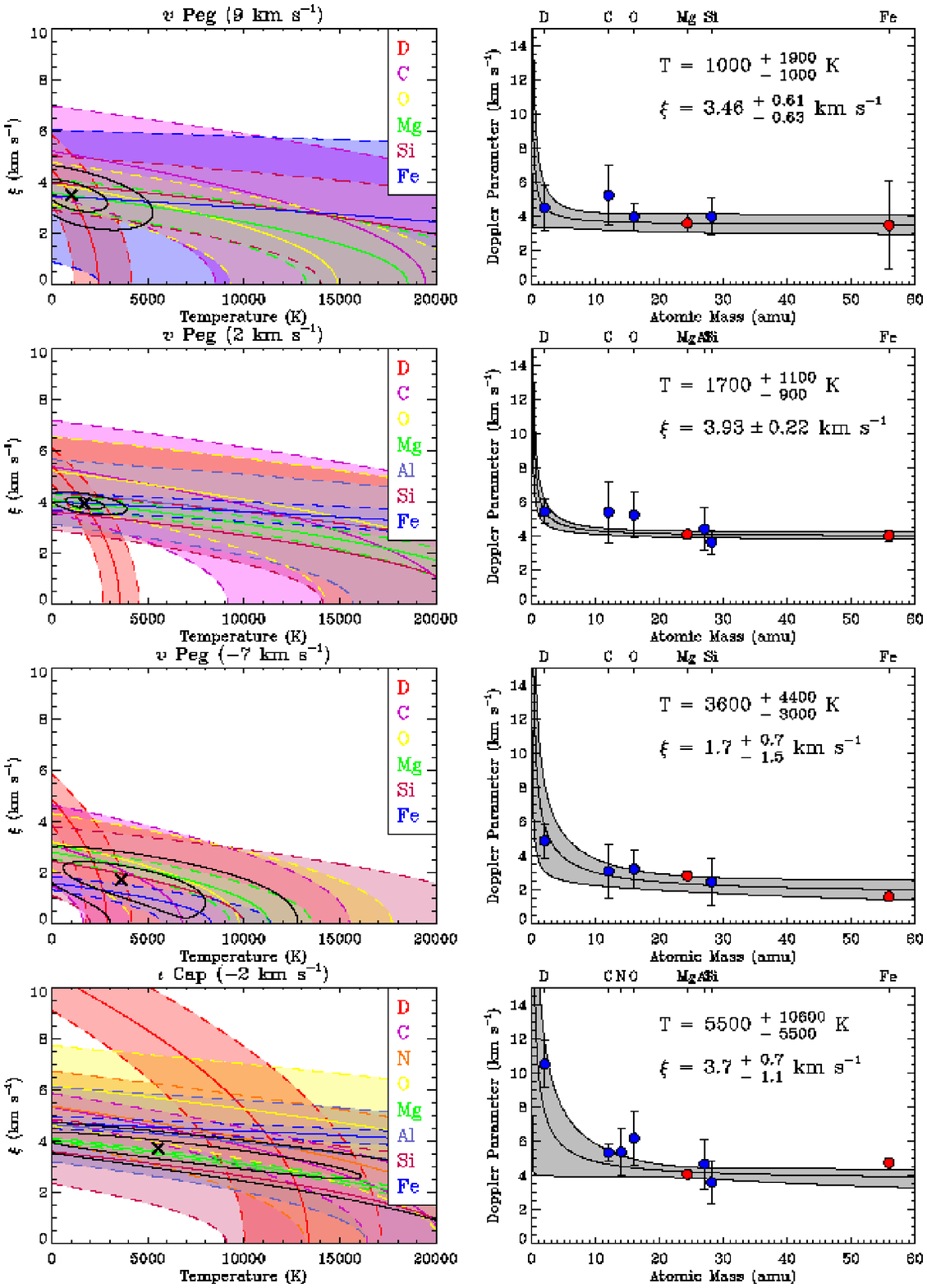}
\caption{Continuation of Figure~\ref{tt_fig1}. \label{tt_fig11}}
\end{figure}

\clearpage
\begin{figure}
\epsscale{.9}
\figurenum{1l}
\plotone{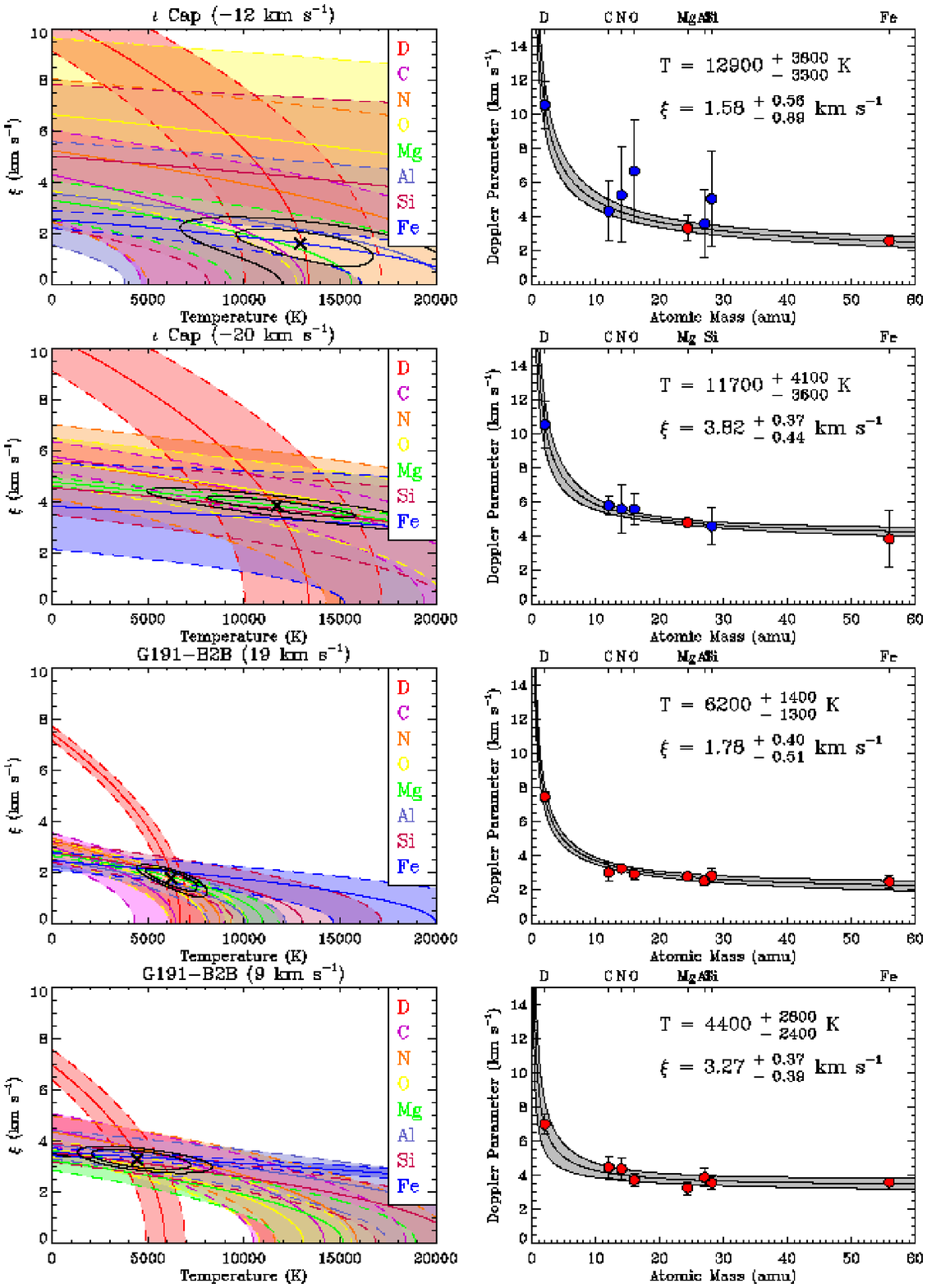}
\caption{Continuation of Figure~\ref{tt_fig1}. \label{tt_fig12}}
\end{figure}

\clearpage
\begin{figure}
\epsscale{.9}
\figurenum{1m}
\plotone{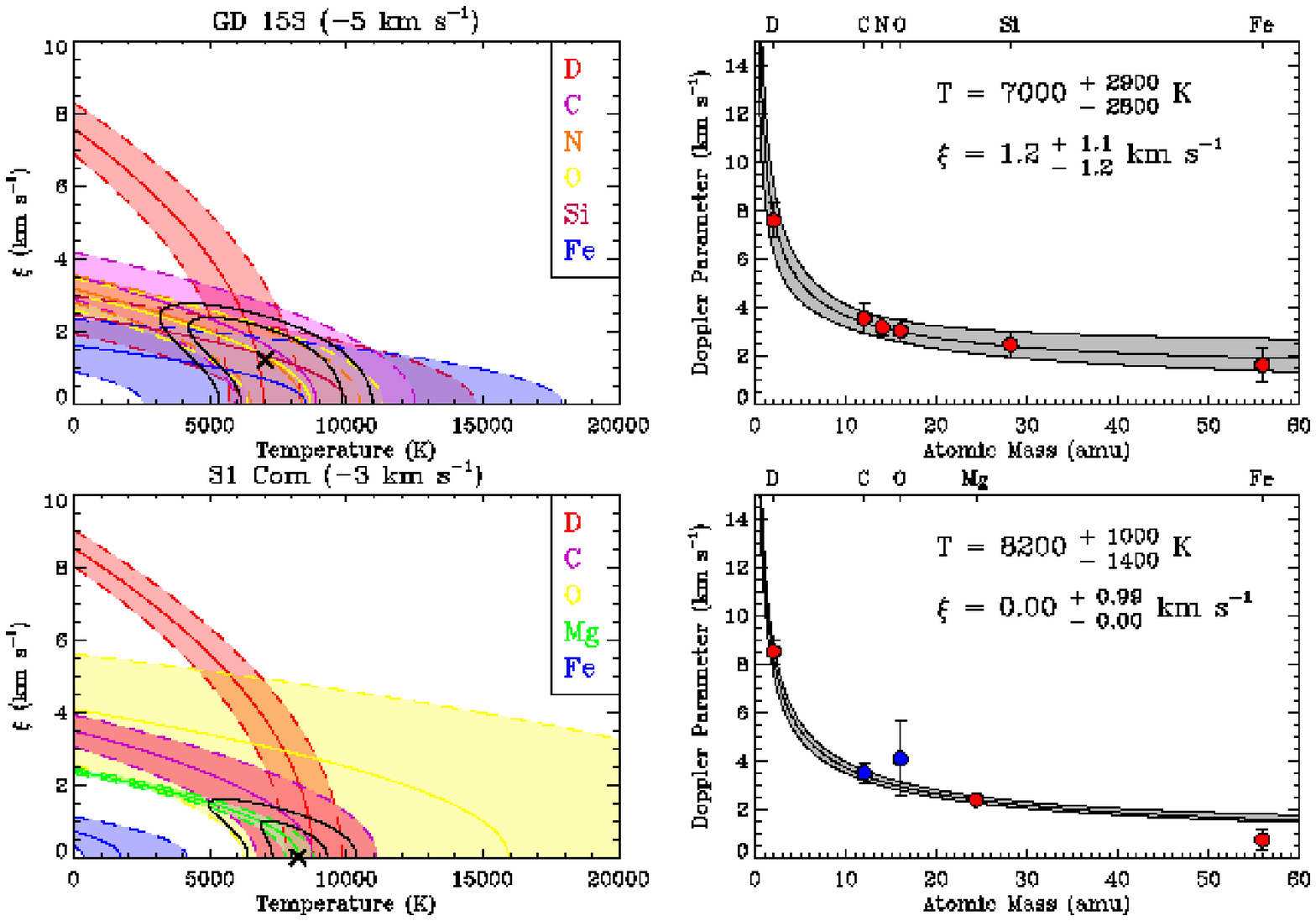}
\caption{Continuation of Figure~\ref{tt_fig1}. \label{tt_fig13}}
\end{figure}

\clearpage
\begin{figure}
\figurenum{2}
\epsscale{1.}
\plotone{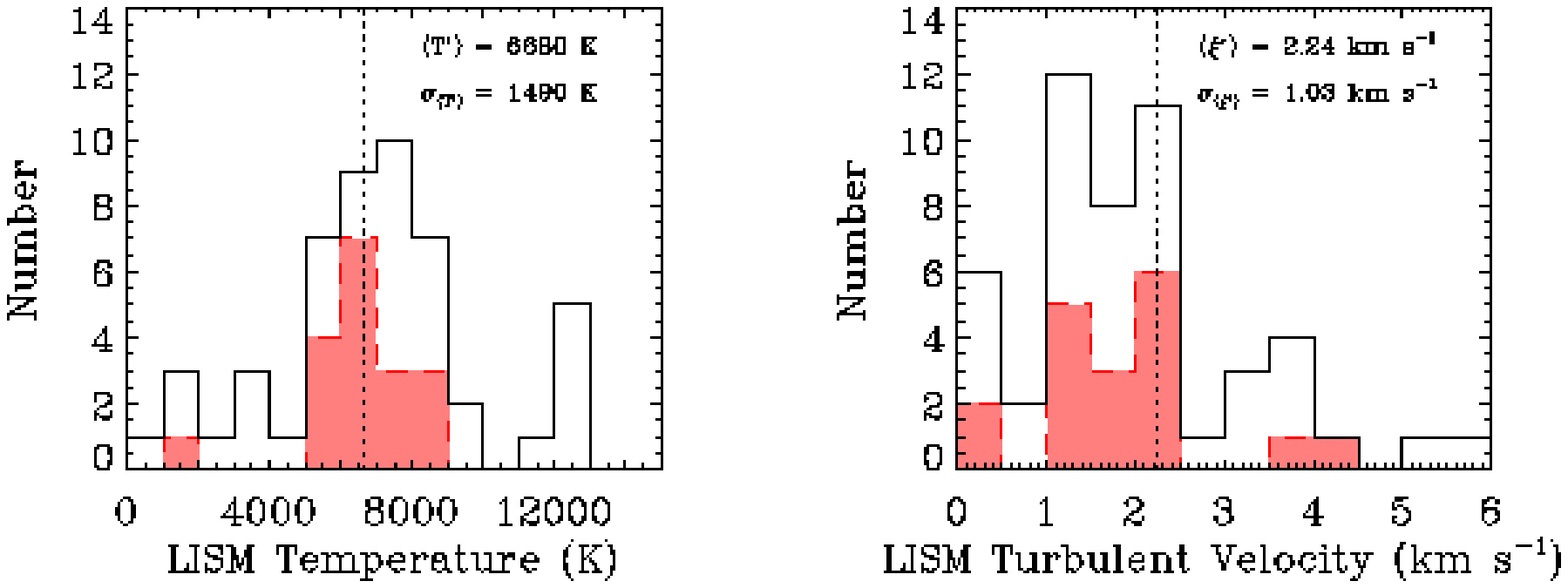}
\caption{Distributions of measured temperatures and turbulent velocities in 
the LISM for clouds within 100\,pc.  The temperature distribution is peaked 
at the canonical LISM temperature $\sim$\,7000\,K.  The weighted mean of our 
sample is $\langle\,T^{\prime}\,\rangle\,=\,$6680\,K, while the dispersion 
about the mean is 1490\,K.  The weighted mean of the turbulent velocity 
distribution is $\langle\,\xi^{\prime}\,\rangle\,=\,$2.24\,km\,s$^{-1}$, with 
a dispersion about the mean of 1.03\,km\,s$^{-1}$.  The weighted mean is 
indicated by a dotted line.  The shaded 
subsamples represent our most precise 
measurements, defined by their absolute error being less than the standard 
deviation of the entire sample.  \label{tt_fig14}}
\end{figure}

\clearpage
\begin{figure}
\figurenum{3}
\epsscale{.56}
\plotone{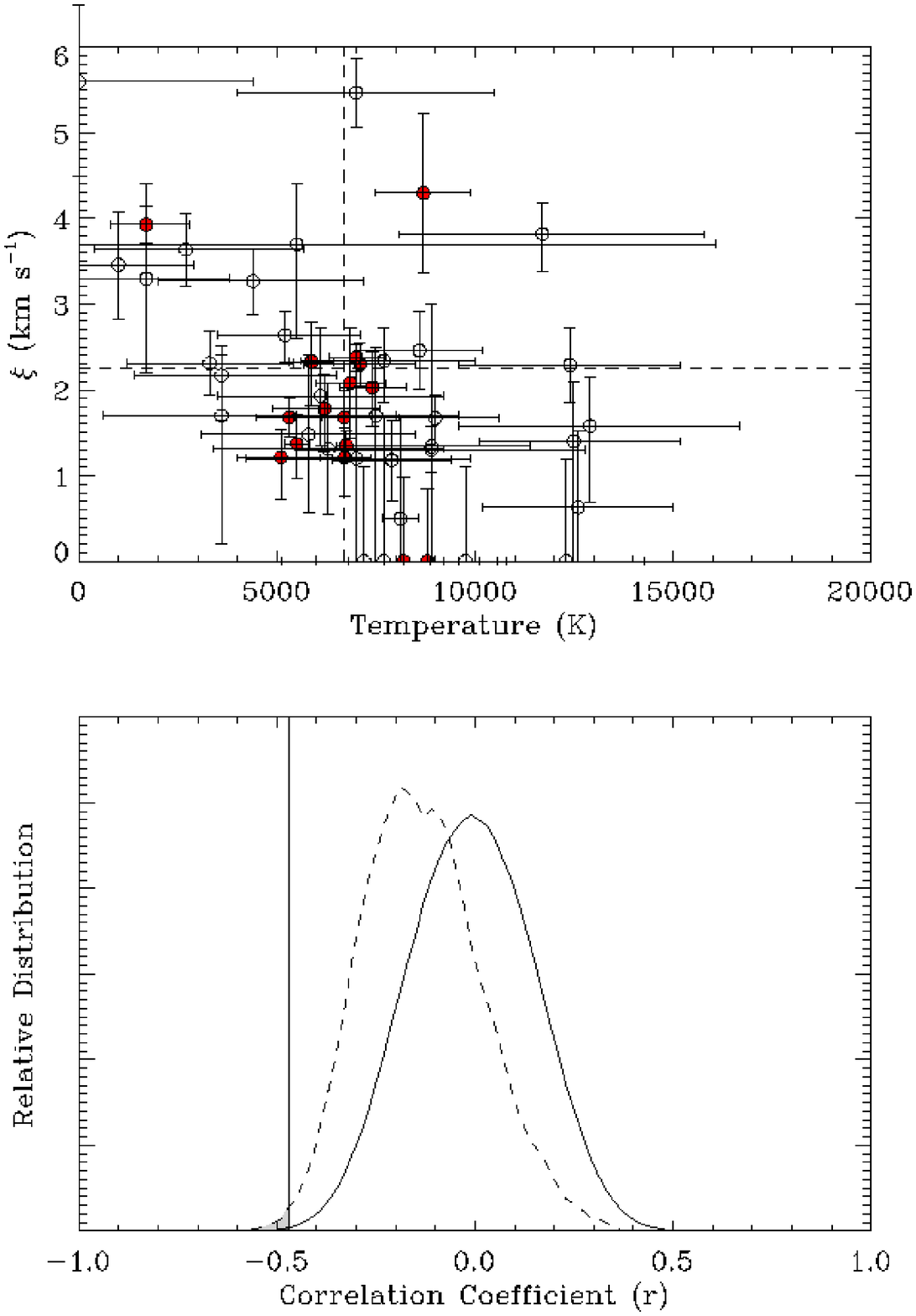}
\caption{The top plot shows the turbulent 
velocity as a function of temperature for our complete 
sample of LISM absorbers within 100\,pc.  The filled symbols indicate our 
most precise measurements, defined by their absolute error being less than 
the standard deviation of the complete sample (see Figure~\ref{tt_fig14}).  
The dashed lines indicate the weighted mean values of the temperature and 
turbulent velocity.  The bottom plot shows the distribution of correlation 
coefficients in $\sim\,10^5$ realizations of two different scenarios.  The 
solid curve shows the distribution of a hypothetical dataset that shows no correlation, varied over the true errors, but assuming the errors in temperature and turbulence are independent and uncorrelated.  The dashed 
curve indicates the same hypothetical sample varied according to the true error bars, but 
 now taking into account the actual 
alignment of the error contours, which indicate a correlation in the errors 
of the variables (see error ellipses in Figure~1).  For all realizations, the 
weighted means and dispersions of both variables ($T$ and $\xi$) were identical to the true sample.  The vertical line indicates the measured correlation of our true sample.  Although the probability that this value of correlation is a result of an uncorrelated parent population increases when we take into account the correlated errors (see shaded regions), the overall probability is still very low ($<\,1$\,\%).  Therefore, 
the covariance between the errors of temperature and turbulence 
do not appear to dominate the distribution of the sample, and 
the moderately negative 
correlation between temperature and turbulence seems to be significant.
\label{tt_fig15}}
\end{figure}

\clearpage
\begin{figure}
\figurenum{4}
\epsscale{.85}
\plotone{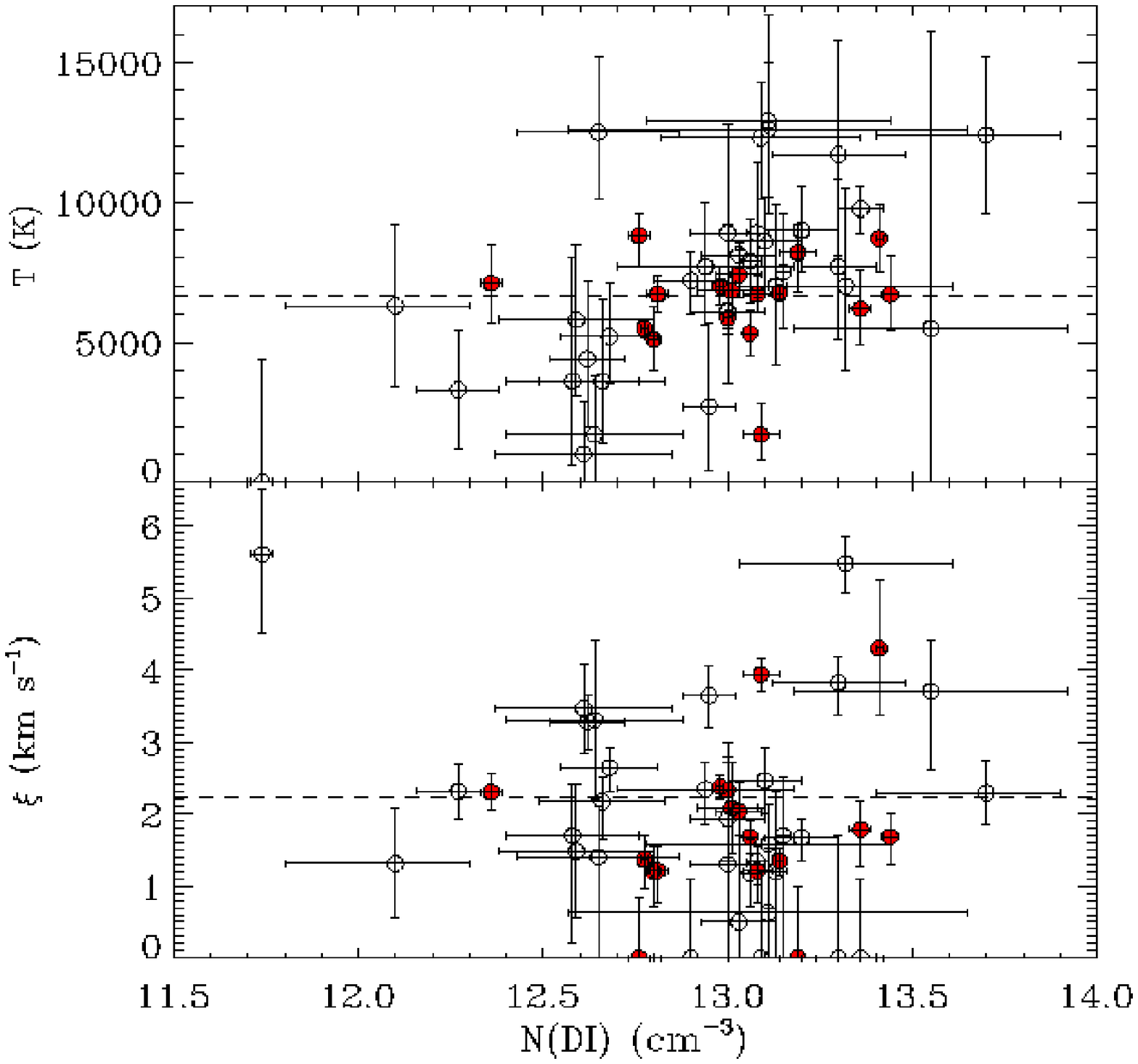}
\caption{Cloud temperature and turbulent velocity 
are plotted against \ion{D}{1} column 
density.  The filled symbols indicate the best measurement subsample.  The 
dashed lines indicate the weighted mean values of the temperature and 
turbulent velocity.  The \ion{D}{1} column density can be converted to a 
\ion{H}{1} column density by applying the locally constant ratio 
D/H$\,=\,(1.56\,\pm\,0.04)\,\times\,10^{-5}$ \citep{wood04}.  The hydrogen 
column density can be used as a proxy for the cloud mass.  \label{ndtxi}}
\end{figure}

\clearpage
\begin{figure}
\figurenum{5}
\epsscale{.75}
\plotone{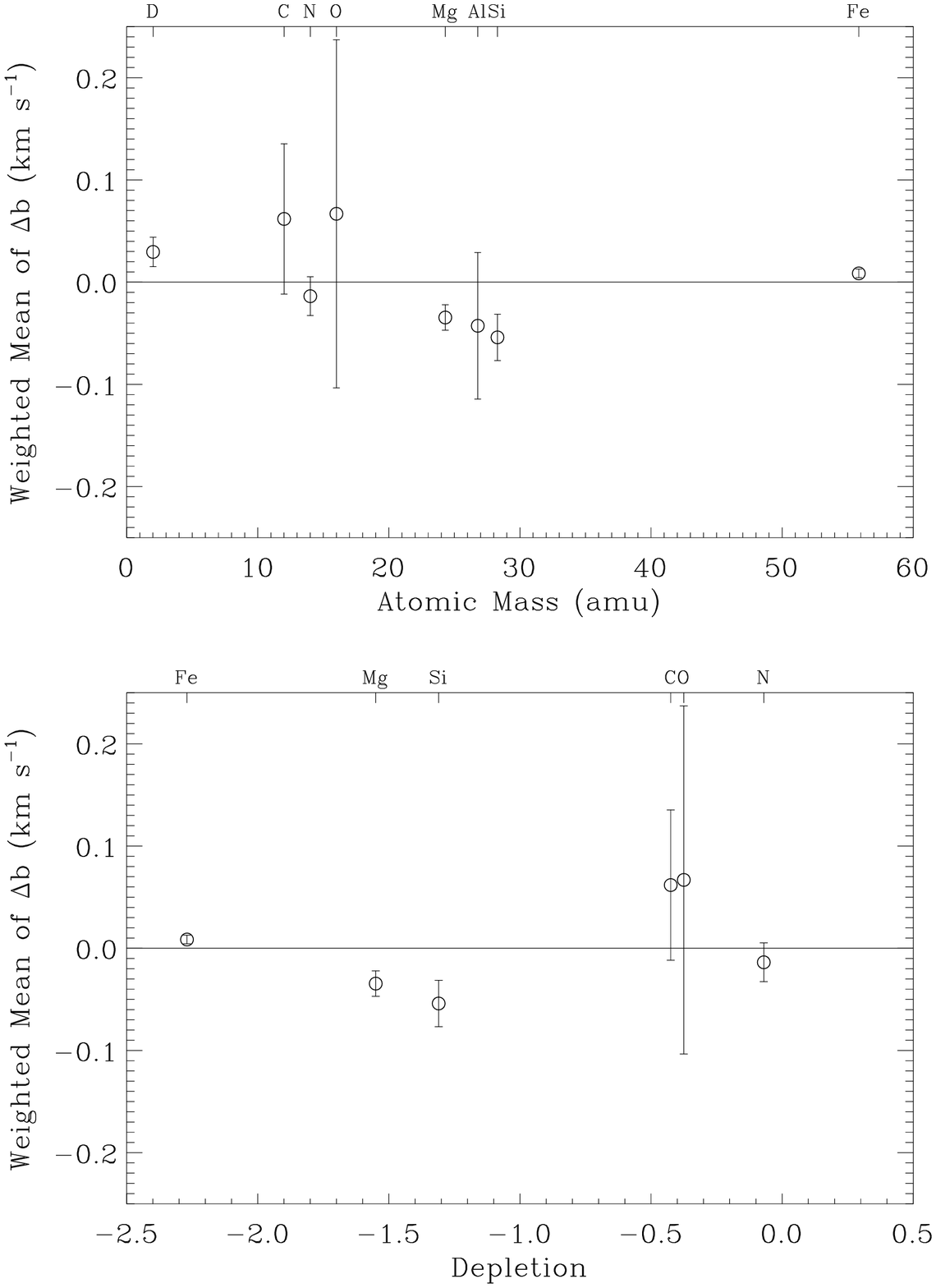}
\caption{Departures of the weighted mean values of the Doppler parameters are 
compared with the best fit values for each individual 
ion.  The top plot shows the weighted mean deviation against atomic mass, and 
the bottom plot shows the same against a typical depletion for each ion in 
the diffuse interstellar medium, taken from the cool diffuse cloud along the $\zeta$~Oph sightline \citep{savage96}.  Although the magnitudes of the depletions of the two diffuse clouds along the line of sight toward $\zeta$~Oph are slightly offset from each other, the general trend displayed here is the same regardless of which cloud is used as the depletion standard..  Carbon and oxygen have 
large deviations 
and uncertainties about the weighted mean because their absorption features 
are typically saturated, which inhibits a precise measurement of the Doppler 
parameter ($b$).  No systematic trends are evident for this class of ions, 
supporting the assumption that they are all tracing the same collection of 
gas in each velocity component.
\label{disper}}
\end{figure}

\clearpage
\begin{figure}
\figurenum{6}
\epsscale{1.}
\plotone{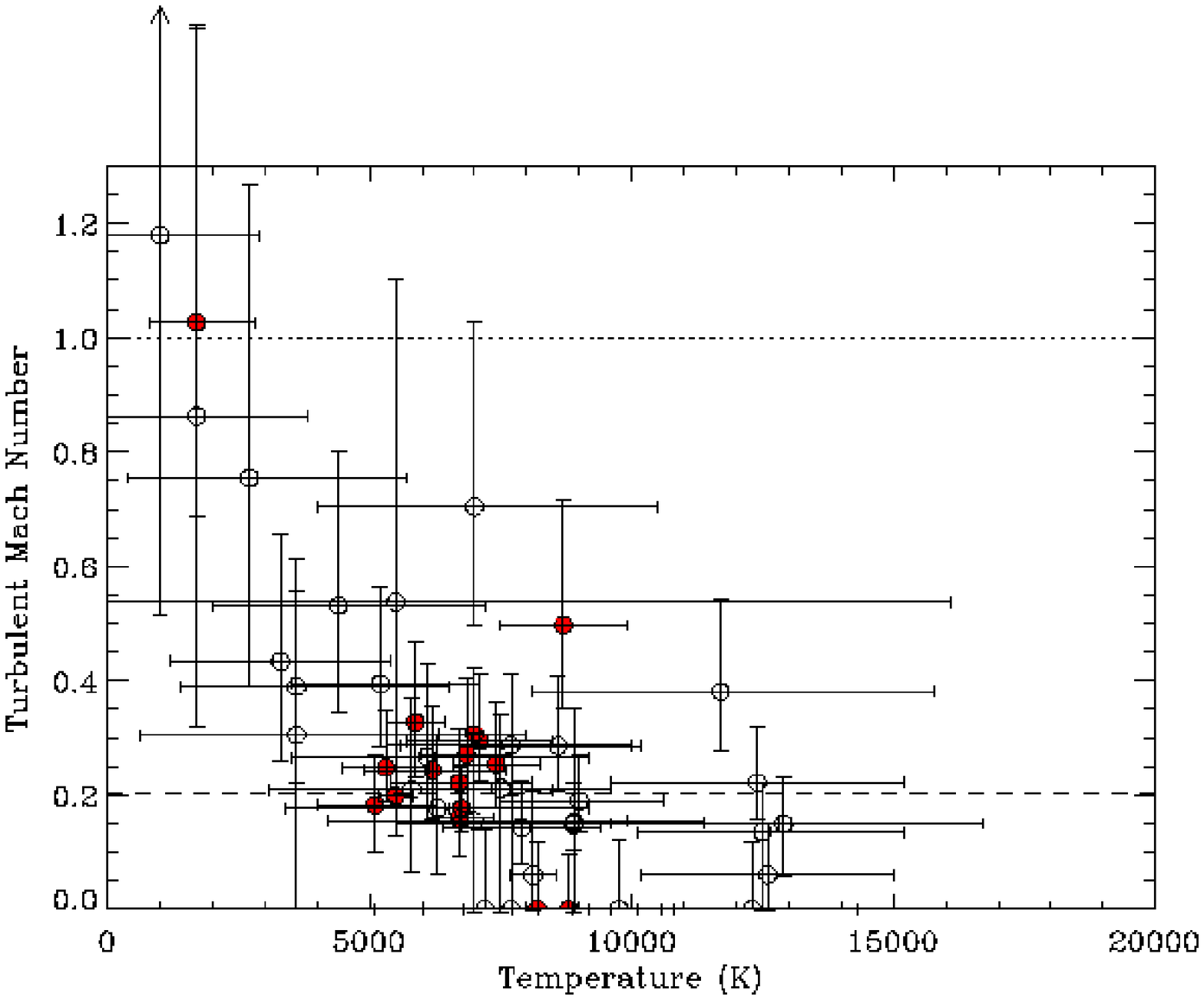}
\caption{A plot of the turbulent Mach number as a 
function of cloud temperature. 
The gradual decrease in turbulent Mach number as a function of 
increasing cloud 
temperature 
is a consequence of the increase in sound speed with temperature, as well as the anticorrelation of turbulent velocity with temperature, discussed in Section~\ref{tt_cor}.  However, the vast 
majority of measurements, 
in particular the best measurement subsample, 
indicate a turbulent Mach number $\sim\,0.2$, shown by the dashed line.  A 
Mach number of unity is indicated by the dotted line.  
\label{tt_f18}}
\end{figure}

\clearpage
\begin{figure}
\figurenum{7}
\epsscale{1.}
\plotone{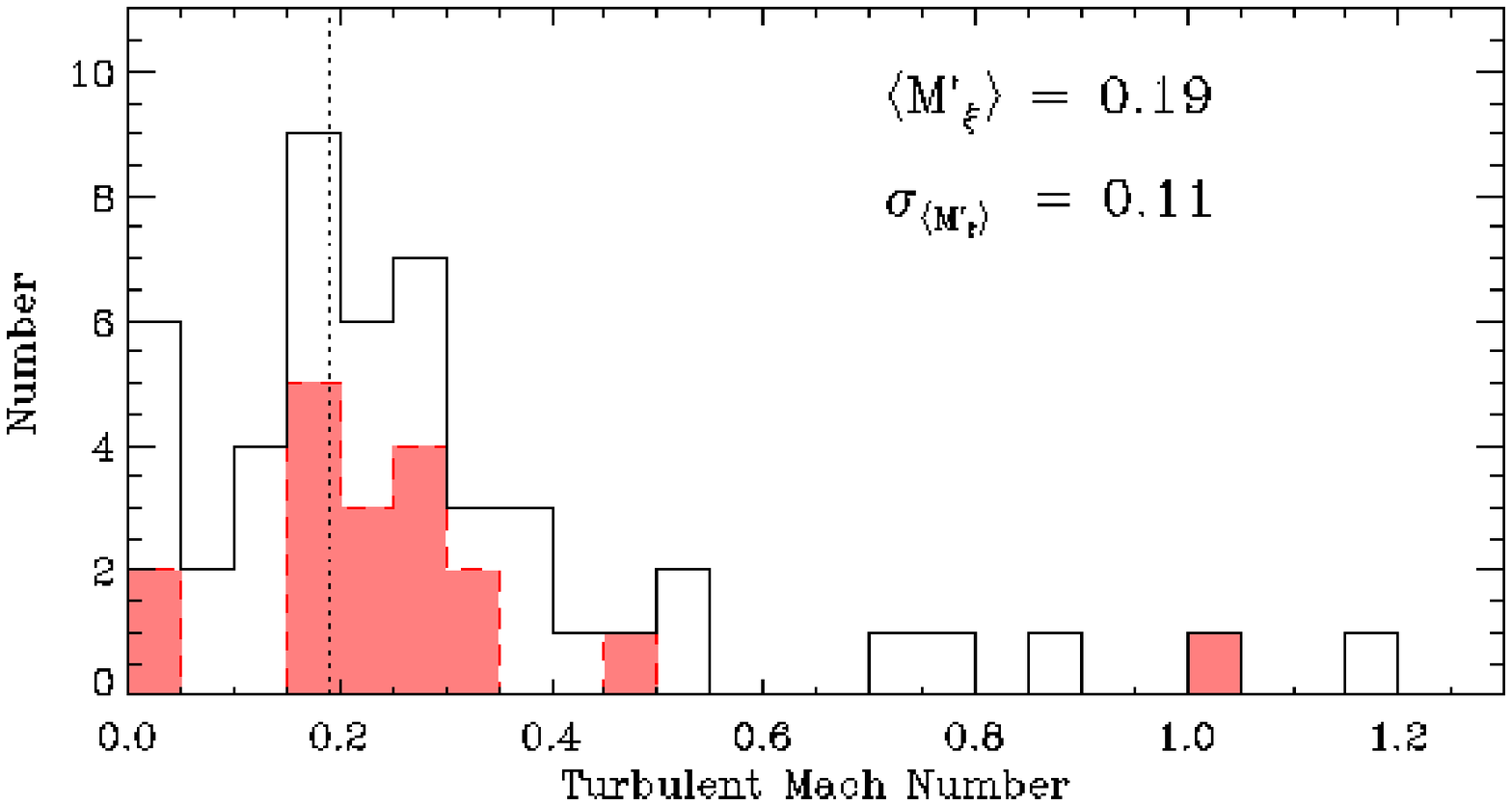}
\caption{Distribution of the observed turbulent Mach numbers in the LISM.  
The weighted mean of our sample is 
$\langle\,M^{\prime}_{\xi}\,\rangle\,=\,$0.19, while the dispersion about the 
mean is 0.11.  The weighted mean is indicated by a dotted line.  The best 
measurement subsample is indicated by the shaded distribution.  
The high values of the turbulent Mach number may be spurious due to the 
possible presence of unresolved closely spaced velocity components. 
Turbulence in clouds within 100\,pc is clearly subsonic.  
\label{tt_f18_5}}
\end{figure}

\clearpage
\begin{figure}
\figurenum{8}
\epsscale{1.}
\plotone{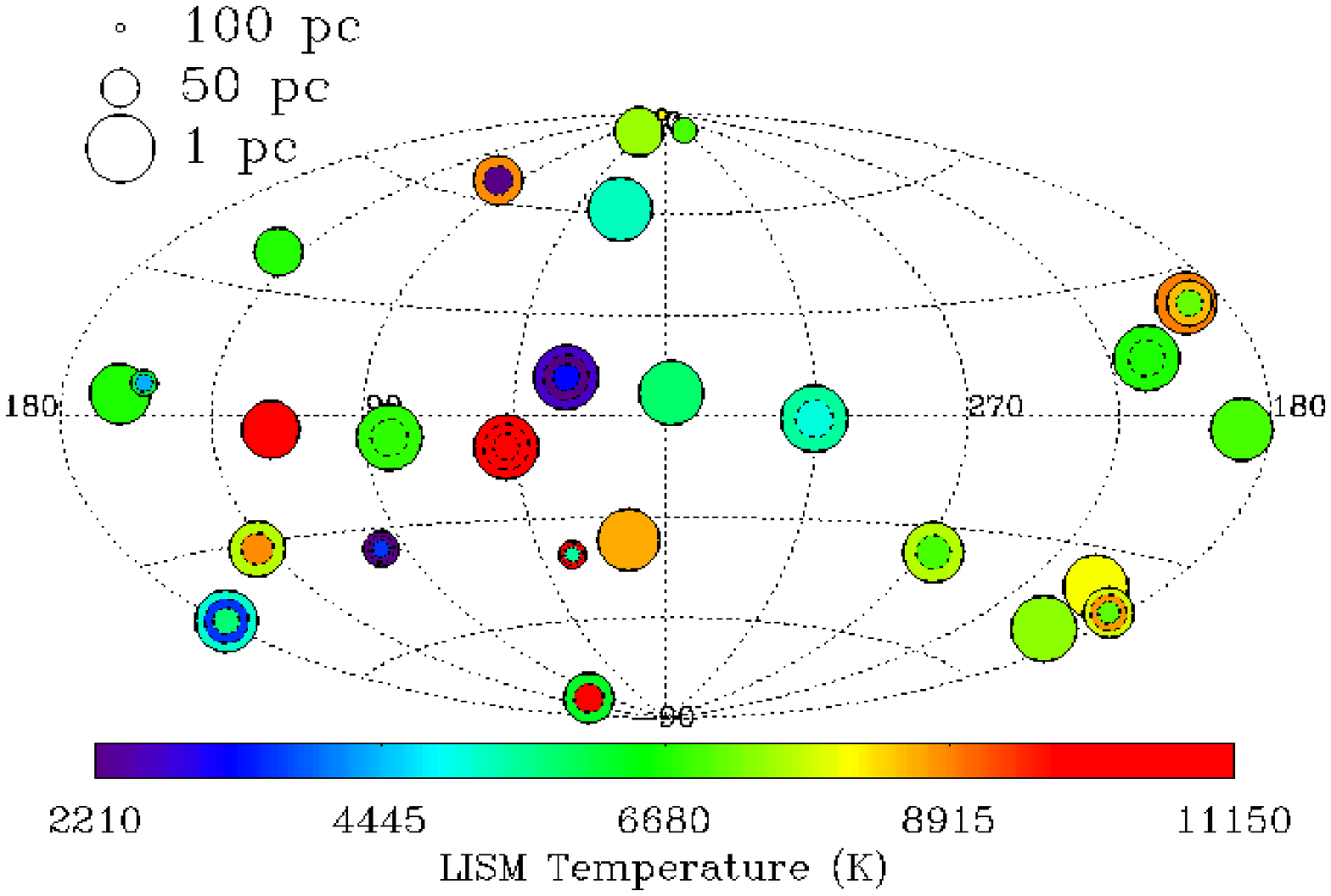}
\caption{The spatial distribution of LISM cloud temperatures is shown in 
Galactic coordinates.  The size of each symbol is inversely proportional to 
the distance, and the 
color of the symbol indicates the temperature of a 
cloud along the line of sight, as given by the scale at the bottom of the 
plot.  If more than one absorbing cloud lies along the line of sight, 
additional cloud temperatures are indicated by nested circles separated by a 
dotted line.  
This technique is also used for measurements of different 
members of a single multiple star system, such $\alpha$~Cen.  
The temperatures for the two velocity components along the lines of sight to 
Procyon and 61 Cyg A are identical because the line widths of the closely spaced velocity components were assumed to be the same.
\label{tt_f19}}
\end{figure}

\clearpage
\begin{figure}
\figurenum{9}
\epsscale{1.}
\plotone{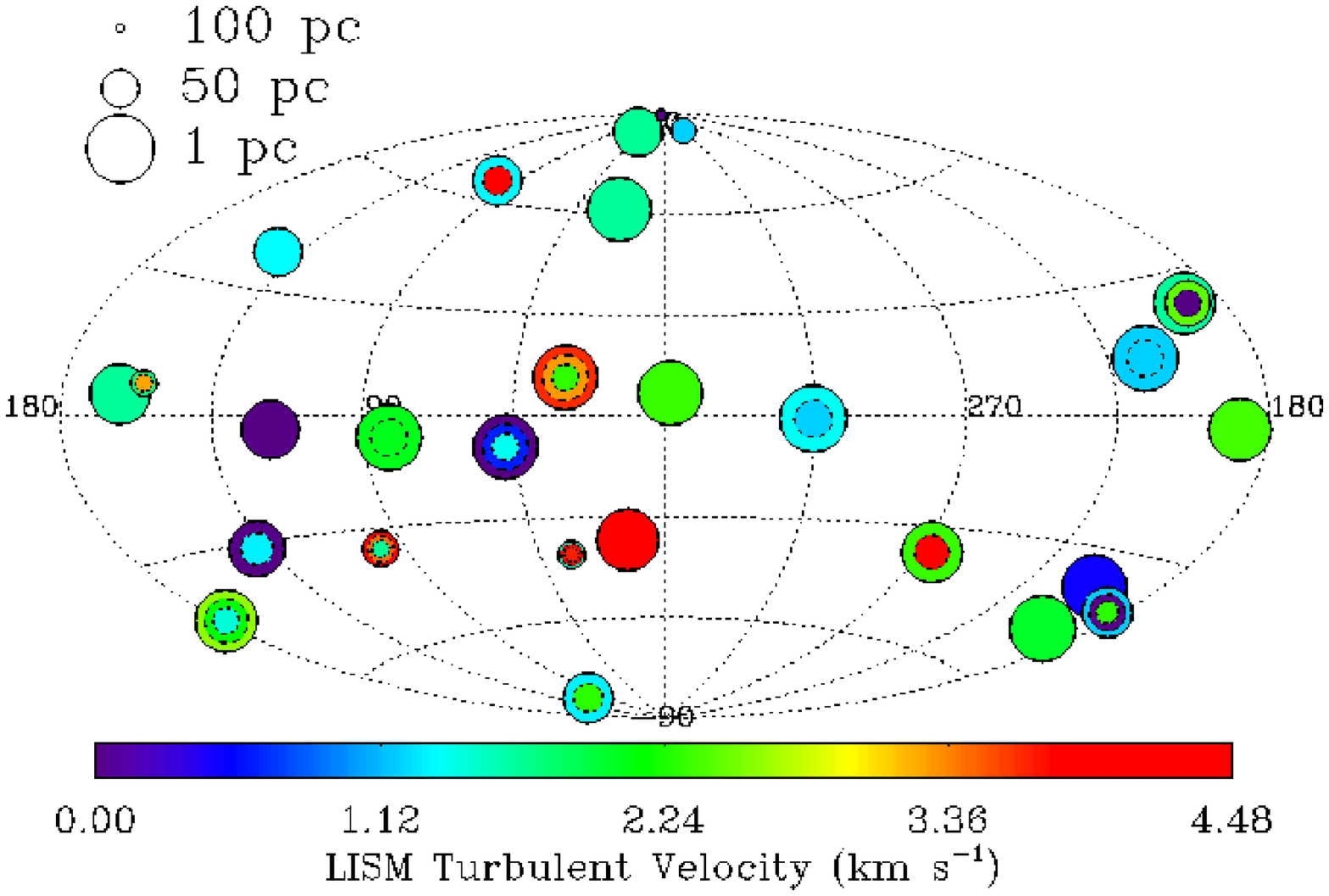}
\caption{Similar to Figure~\ref{tt_f19}, but displaying the turbulent 
velocity of LISM material within 100\,pc.  
\label{tt_f20}}
\end{figure}

\clearpage
\begin{figure}
\figurenum{10}
\epsscale{.7}
\plotone{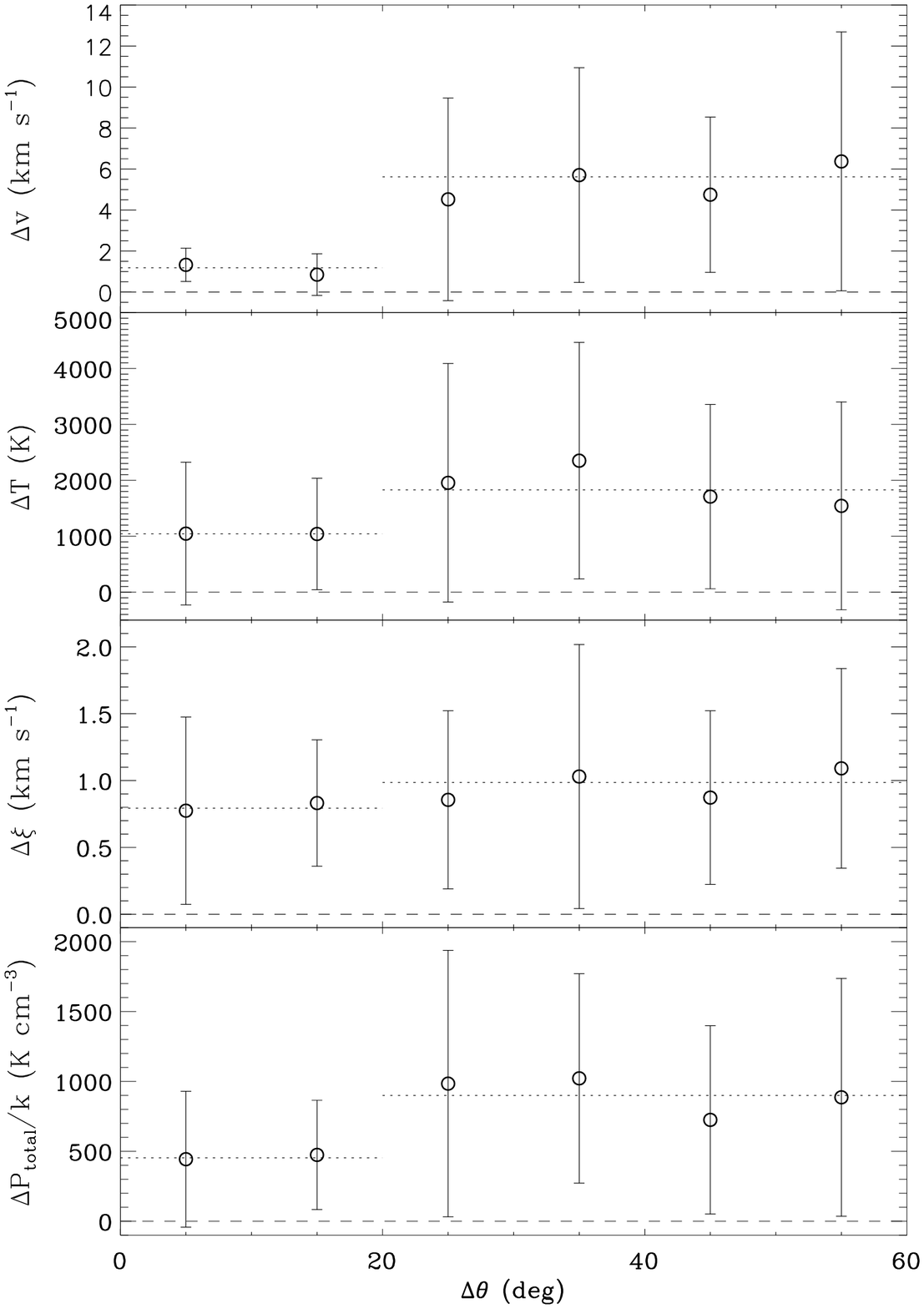}
\caption{Differences between the measured physical parameters, velocity, 
temperature, turbulent velocity, and total pressure, as a function of angular 
separation.  The bin size is 10$^{\circ}$, and the weighted mean for each bin 
is depicted by the circle symbol, while the dispersion about the weighted 
mean is 
indicated by the error bars.  On average, 20 pairings are used in each 
bin, while the minimum, in the 10$^{\circ}$-20$^{\circ}$ bin, is 6 pairings.  
A spatial correlation is obvious for the velocity measurements, where the 
absolute velocity differences, and the magnitude of the dispersion about the 
mean are much smaller for pairs of sightlines that are separated by 
20$^{\circ}$ or less.  The weighted mean for separations from 
0$^{\circ}$-20$^{\circ}$ and 20$^{\circ}$-60$^{\circ}$ are indicated in each 
plot by the dotted line.  A similar spatial correlation seems to be evident 
in the total pressure plot, and to a lesser degree in the individual 
temperature and turbulent velocity plots.
\label{tt_f20_5}}
\end{figure}

\clearpage
\begin{deluxetable}{llcccclll}
\tablewidth{0pt}
\tabletypesize{\tiny}
\tablecaption{Individual Temperature and Turbulent Velocity Measurements
\label{tt_table1}}
\tablehead{ HD & Other & $l$ & $b$ & distance & $\langle v \rangle$ & 
\multicolumn{1}{c}{$T$} & \multicolumn{1}{c}{$\xi$} & Ions used \\
\# & Name & ($^{\circ}$) & ($^{\circ}$) & (pc) & (km s$^{-1}$) & 
\multicolumn{1}{c}{(K)} & \multicolumn{1}{c}{(km s$^{-1}$)}& \\}
\startdata
128620 &$\alpha$ Cen A& 315.7 & --0.7  & 1.3  & --18.45 $\pm$ 0.32 & 
\phn5100$\:$\,$^{+\,1200}_{-\,1100}$ & 1.21$\:$\,$^{+\,0.33}_{-\,0.49}$ & 
DI, CII, NI, OI, MgII, AlII, SiII, FeII \\
128621 &$\alpha$ Cen B& 315.7 & --0.7  & 1.3  & --18.14 $\pm$ 0.08 & 
\phn5500$\:$\,$^{+\,330}_{-\,320}$ & 1.37$\:$\,$^{+\,0.34}_{-\,0.41}$ & 
DI, MgII, FeII \\
22049  &$\epsilon$ Eri& 195.8 & --48.1 & 3.2  & \php18.73 $\pm$ 0.65 & 
\phn7410$\:$\,$^{+\,860}_{-\,830}$ & 2.03$\:$\,$^{+\,0.41}_{-\,0.45}$ & 
DI, CII, OI, MgII, SiII, FeII \\
201091 &61 Cyg A& 82.3  & --5.8  & 3.5  & \phn--3.0 $\pm$ 1.4 & 
\phn6850 $\pm$ 880 & 2.08 $\pm$ 0.64 & 
DI, MgII \\
       &        & & & & \phn--9.0 $\pm$ 0.7 & 
\phn6850 $\pm$ 880 & 2.08 $\pm$ 0.64 & 
DI, MgII \\
61421  &Procyon & 213.7 & 13.0   & 3.5  & \phn23.0 $\pm$ 1.0 & 
\phn6710$\:$\,$^{+\,660}_{-\,630}$ & 1.21$\:$\,$^{+\,0.35}_{-\,0.45}$ & 
DI, MgII, FeII \\
       &        & & & & \phn19.76 $\pm$ 0.75 & 
\phn6710$\:$\,$^{+\,660}_{-\,630}$ & 1.21$\:$\,$^{+\,0.35}_{-\,0.45}$ & 
DI, MgII, FeII \\
26965  &40 Eri A& 200.8 & --38.1 & 5.0  & \phn21.73 $\pm$ 0.22 & 
\phn8120 $\pm$ 450 & 0.5\phn$\:$\,$^{+\,1.2}_{-\,0.5}$ & 
DI, MgII \\
165341 &70 Oph  & 29.9  & 11.4   & 5.1  & --26.50 $\pm$ 0.06 & 
\phn2700$\:$\,$^{+\,3000}_{-\,2300}$ & 3.64$\:$\,$^{+\,0.42}_{-\,0.44}$ & 
DI, CII, OI, MgII, SiII, FeII \\
       &        & & & & --32.53 $\pm$ 0.32 & 
\phn1700$\:$\,$^{+\,2100}_{-\,1700}$ & 3.3 $\pm$ 1.1 & 
DI, CII, OI, MgII, SiII \\
       &        & & & & --43.34 $\pm$ 0.14 & 
\phn3300 $\pm$ 2100 & 2.31 $\pm$ 0.37 & 
DI, CII, OI, MgII, SiII \\
187642 &Altair  & 47.7  & --8.9  & 5.1  & --17.1 $\pm$ 1.2 & 
12300$\:$\,$^{+\,2000}_{-\,2200}$ & 0.0\phn$\:$\,$^{+\,1.2}_{-\,0.0}$ & 
DI, CII, MgII, FeII \\
       &        & & & & --20.9 $\pm$ 1.2 & 
12600 $\pm$ 2400 & 0.63\phn$\:$\,$^{+\,0.90}_{-\,0.63}$ & 
DI, CII, MgII, FeII \\
       &        & & & & --25.02 $\pm$ 0.84 & 
12500$\:$\,$^{+\,2700}_{-\,2400}$ & 1.4\phn$\:$\,$^{+\,0.7}_{-\,1.4}$ & 
DI, CII, MgII, FeII \\
155886 &36 Oph A& 358.3 & 6.9    & 5.5  & --28.40 $\pm$ 0.58 & 
\phn5870 $\pm$ 560 & 2.33$\:$\,$^{+\,0.46}_{-\,0.51}$ & 
DI, MgII, FeII \\
131156A&$\xi$ Boo A& 23.1  & 61.4   & 6.7  & --17.69 $\pm$ 0.60 & 
\phn5310 $\pm$ 830 & 1.68 $\pm$ 0.23 & 
DI, CII, OI, MgII, SiII \\
39587  &$\chi^1$ Ori& 188.5 & --2.7  & 8.7  & \phn23.08 $\pm$ 0.66 & 
\phn7000$\:$\,$^{+\,730}_{-\,680}$ & 2.38$\:$\,$^{+\,0.15}_{-\,0.17}$ & 
DI, CII, MgII, SiII, FeII \\
20630  &$\kappa^1$ Cet& 178.2 & --43.1 & 9.2  & \phn20.84 $\pm$ 0.20 & 
\phn5200$\:$\,$^{+\,1900}_{-\,1700}$ & 2.64$\:$\,$^{+\,0.28}_{-\,0.32}$ & 
DI, CII, OI, MgII, SiII, FeII \\
       &              & & & & \phn13.35 $\pm$ 0.15 & 
\phn3600$\:$\,$^{+\,2900}_{-\,2200}$ & 2.17$\:$\,$^{+\,0.34}_{-\,0.52}$ & 
DI, CII, OI, MgII, SiII, FeII \\
       &              & & & & \phn\phn7.36 $\pm$ 0.46 & 
\phn5800 $\pm$ 2700 & 1.48 $\pm$ 0.92 & 
DI, CII, OI, MgII, SiII \\
197481 & AU Mic & 12.7 & --36.8 & 9.9  & --21.45 $\pm$ 0.34 & 
\phn8700 $\pm$ 1200 & 4.30 $\pm$ 0.93 & 
DI, MgII \\
62509  &$\beta$ Gem& 192.2 & 23.4   & 10.3 & \phn31.84 $\pm$ 0.60 & 
\phn6100$\:$\,$^{+\,3100}_{-\,2600}$ & 1.93$\:$\,$^{+\,0.79}_{-\,0.59}$ & 
DI, OI, MgII, FeII \\
       &           & & & & \phn19.65 $\pm$ 0.76 & 
\phn9000$\:$\,$^{+\,1600}_{-\,1500}$ & 1.67$\:$\,$^{+\,0.27}_{-\,0.32}$ & 
DI, OI, MgII, FeII \\
33262  &$\zeta$ Dor& 266.0 & --36.7 & 11.7 & \phn13.9 $\pm$ 1.3 & 
\phn7000$\:$\,$^{+\,3500}_{-\,3000}$ & 5.47$\:$\,$^{+\,0.39}_{-\,0.41}$ & 
DI, CII, OI, MgII, AlII, SiII, FeII \\
       &           & & & & \phn\phn8.41 $\pm$ 0.82 & 
\phn7700$\:$\,$^{+\,2300}_{-\,2100}$ & 2.34$\:$\,$^{+\,0.38}_{-\,0.48}$ & 
DI, CII, OI, MgII, AlII, SiII, FeII \\
34029  &Capella & 162.6 & 4.6    & 12.9 & \phn21.48 $\pm$ 0.51 & 
\phn6700$\:$\,$^{+\,1400}_{-\,1300}$ & 1.68$\:$\,$^{+\,0.32}_{-\,0.39}$ & 
DI, CII, NI, OI, MgII, AlII, SiII, FeII \\
432    &$\beta$ Cas& 117.5 & --3.3  & 16.7 & \phn\phn9.15 $\pm$ 0.68 & 
\phn9760$\:$\,$^{+\,800}_{-\,880}$ & 0.0\phn$\:$\,$^{+\,1.1}_{-\,0.0}$ & 
DI, MgII, FeII \\
11443  &$\alpha$ Tri& 138.6 & --31.4 & 19.7 & \phn17.89 $\pm$ 0.57 & 
\phn7700$\:$\,$^{+\,3100}_{-\,2600}$ & 0.0\phn$\:$\,$^{+\,1.7}_{-\,0.0}$ & 
DI, MgII, FeII \\
       &            & & & & \phn13.65 $\pm$ 0.75 & 
\phn8900$\:$\,$^{+\,3900}_{-\,3400}$ & 1.3\phn$\:$\,$^{+\,1.7}_{-\,1.3}$ & 
DI, MgII, FeII \\
22468  &HR 1099& 184.9 & --41.6 & 29.0 & \phn21.90 $\pm$ 0.04 & 
\phn7900 $\pm$ 1500 & 1.18 $\pm$ 0.47 & 
HI, CII, OI, MgII \\
       &        & & & & \phn14.80 $\pm$ 0.02 & 
\phn8800$\:$\,$^{+\,800}_{-\,1100}$ & 0.00$\:$\,$^{+\,0.85}_{-\,0.00}$ & 
HI, CII, OI, MgII \\
       &        & & & & \phn\phn8.20 $\pm$ 0.02 & 
\phn7100 $\pm$ 1400 & 2.30 $\pm$ 0.25 & 
HI, CII, OI, MgII \\
4128   &$\beta$ Cet& 111.3 & --80.7 & 29.4 & \phn\phn9.14 $\pm$ 0.21 & 
\phn6300 $\pm$ 2900 & 1.31 $\pm$ 0.76 & 
HI, CII, NI, OI, MgII, AlII, SiII \\
       &           & & & & \phn\phn1.63 $\pm$ 0.22 & 
12400 $\pm$ 2800 & 2.29 $\pm$ 0.44 & 
HI, CII, NI, OI, MgII, AlII, SiII \\
120315 &Alcaid  & 100.7 & 65.3 & 30.9 & \phn\phn2.6 $\pm$ 3.4 & 
\phn\phn\phn\phn0$\:$\,$^{+\,4400}_{-\,0}$ & 5.6$\:$\,$^{+\,0.9}_{-\,1.1}$ & 
DI, CII, MgII, SiII \\
       &        & & & & \phn--3.02 $\pm$ 0.51 & 
\phn8900$\:$\,$^{+\,2500}_{-\,2300}$ & 1.34$\:$\,$^{+\,0.24}_{-\,0.31}$ & 
DI, CII, OI, MgII, SiII, FeII \\
       &HZ 43   & 54.1  & 84.2   & 32.0 & \phn--6.52 $\pm$ 0.39 & 
\phn7500$\:$\,$^{+\,2100}_{-\,2000}$ & 1.7\phn$\:$\,$^{+\,0.8}_{-\,1.7}$ & 
DI, CII, NI, OI, MgII, SiII, FeII \\
82210  &DK UMa  & 142.6 & 38.9   & 32.4 & \phn\phn9.41 $\pm$ 0.61 & 
\phn6750 $\pm$ 240 & 1.35$\:$\,$^{+\,0.18}_{-\,0.20}$ & 
DI, CII, OI, MgII, AlII, FeII \\
62044  &$\sigma$ Gem& 191.2 & 23.3   & 37.5 & \phn32.26 $\pm$ 0.15 & 
\phn7200$\:$\,$^{+\,1000}_{-\,1200}$ & 0.0\phn$\:$\,$^{+\,1.1}_{-\,0.0}$ & 
DI, MgII \\
       &            & & & & \phn21.77 $\pm$ 0.16 & 
\phn8600 $\pm$ 1600 & 2.46 $\pm$ 0.45 & 
DI, MgII \\
220657 &$\upsilon$ Peg& 98.6  & --35.4 & 53.1 & \phn\phn8.8 $\pm$ 1.2 & 
\phn\phn1000$\:$\,$^{+\,1900}_{-\,1000}$ & 3.46$\:$\,$^{+\,0.61}_{-\,0.63}$ & 
DI, CII, OI, MgII, SiII, FeII \\
       &              & & & & \phn\phn1.73 $\pm$ 0.39 & 
\phn1700$\:$\,$^{+\,1100}_{-\,900}$ & 3.93 $\pm$ 0.22 & 
DI, CII, OI, MgII, AlII, SiII, FeII \\
       &              & & & & \phn--7.48 $\pm$ 0.43 & 
\phn3600$\:$\,$^{+\,4400}_{-\,3000}$ & 1.7$\:$\,$^{+\,0.7}_{-\,1.5}$ & 
DI, CII, OI, MgII, SiII, FeII \\
203387 &$\iota$ Cap& 33.6  & --40.8 & 66.1 & \phn--2.22 $\pm$ 0.19 & 
\phn5500$\:$\,$^{+\,10600}_{-\,5500}$ & 3.7$\:$\,$^{+\,0.7}_{-\,1.1}$ & 
DI, CII, NI, OI, MgII, AlII, SiII, FeII \\
       &           & & & & --12.06 $\pm$ 0.49 & 
12900$\:$\,$^{+\,3800}_{-\,3300}$ & 1.58\phn$\:$\,$^{+\,0.56}_{-\,0.89}$ & 
DI, CII, NI, OI, MgII, AlII, SiII, FeII \\
       &           & & & & --20.48 $\pm$ 0.45 & 
11700$\:$\,$^{+\,4100}_{-\,3600}$ & 3.82$\:$\,$^{+\,0.37}_{-\,0.44}$ & 
DI, CII, NI, OI, MgII, SiII, FeII \\
     & G191-B2B & 156.0 & 7.1   & 68.8 & \phn19.19 $\pm$ 0.15& 
\phn6200$\:$\,$^{+\,1400}_{-\,1300}$ & 1.78$\:$\,$^{+\,0.40}_{-\,0.51}$ & 
DI, CII, NI, OI, MgII, AlII, SiII, FeII \\
     &          & & & & \phn\phn8.61 $\pm$ 0.38 & 
\phn4400$\:$\,$^{+\,2800}_{-\,2400}$ & 3.27$\:$\,$^{+\,0.37}_{-\,0.39}$ & 
DI, CII, NI, OI, MgII, AlII, SiII, FeII \\
       & GD 153 & 317.3 & 84.8 & 70.5 & \phn--5.04 $\pm$ 0.17 & 
\phn7000$\:$\,$^{+\,2900}_{-\,2800}$ & 1.2\phn$\:$\,$^{+\,1.1}_{-\,1.2}$ & 
DI, CII, NI, OI, SiII, FeII \\
111812 &31 Com  & 115.0 & 89.6 & 94.2 & \phn--3.37 $\pm$ 0.37 & 
\phn8200$\:$\,$^{+\,1000}_{-\,1400}$ & 0.00$\:$\,$^{+\,0.99}_{-\,0.00}$ & 
DI, CII, OI, MgII, FeII \\

\enddata
\tablecomments{This table includes only those measurements that include 
independent fits to the line width of deuterium and an ion at least as heavy 
as magnesium.}
\end{deluxetable}

\end{document}